\documentclass[ALICE,manyauthors]{cernphprep}
\usepackage{AliJetv2pPb}
\usepackage[T1]{fontenc}
\usepackage{orcidlink}

\DeclareUnicodeCharacter{00A0}{}

\begin{document}

\begin{titlepage}
\PHyear{2022}
\PHnumber{290}   
\PHdate{20 December}  

\title{Azimuthal anisotropy of jet particles in \\
p--Pb and Pb--Pb collisions at $\mathbf{\sqrt{{\textit s}_{\rm\mathbf {NN}}}}$ = $\mathbf{5.02}$~TeV
}
\ShortTitle{Azimuthal anisotropy of jet particles in p--Pb and Pb--Pb collisions}  

\Collaboration{ALICE Collaboration\thanks{See Appendix~\ref{app:collab} for the list of collaboration members}}
\ShortAuthor{ALICE Collaboration} 

\begin{abstract}
  The azimuthal anisotropy of particles associated with jets (jet particles) at midrapidity is measured for the first time in \pPb and \PbPb collisions at \snn $=$ 5.02 TeV down to transverse momentum (\pt) of 0.5~GeV/$c$ and 2~GeV/$c$, respectively, with ALICE. The results obtained in \pPb collisions are based on a novel three-particle correlation technique. The azimuthal anisotropy coefficient $v_2$ in high-multiplicity \pPb collisions is positive, with a significance reaching 6.8$\sigma$ at low \pt, and its magnitude is smaller than in semicentral \PbPb collisions. In contrast to the measurements in \PbPb collisions, the $v_2$ coefficient is also found independent of \pt within uncertainties. Comparisons with the inclusive charged-particle $v_2$ and with AMPT calculations are discussed. The predictions suggest that parton interactions play an important role in generating a non-zero jet-particle $v_2$ in \pPb collisions, even though they overestimate the reported measurement. These observations shed new insights on the understanding of the origin of the collective behaviour of jet particles in small systems such as \pPb collisions, and provide significant stringent new constraints to models. 

\end{abstract}
\end{titlepage}

\setcounter{page}{2}


\section{Introduction}\label{sec:intro}

The study of ultrarelativistic heavy-ion collisions aims to investigate the properties of strongly-interacting matter characterised by high energy density and temperature, known as the quark--gluon plasma (QGP)~\cite{Bazavov:2014pvz,Shuryak:2014zxa}. In non-central collisions, the initial spatial anisotropy of the overlap region of the colliding nuclei is converted into an anisotropy in momentum space via interactions among the medium constituents. The final-state anisotropies are quantified by the coefficients ($v_n$) of a Fourier decomposition of the azimuthal ($\varphi$) distribution of produced particles~\cite{Voloshin:1994mz,Poskanzer:1998yz}:
\begin{equation}
 {{\rm d}^2N \over {{\rm d}p_{\rm T}{\rm d}\varphi }} = {1\over 2\pi} 
{{\rm d}N \over {\rm d} p_{\rm T}} 
\bigg ( 1 + 2 \sum_{{n} = 1}^{\infty} v_{n} (p_{\rm T}) {\rm cos} 
\lbrack {n} ( \varphi - \Psi_{n}) \rbrack \bigg),
\label{eq:phi}
\end{equation}
where \pt is the transverse momentum and $\Psi_{n}$ is the azimuthal angle of the symmetry plane for the $n^{\rm th}$ harmonic. Measurements of the $v_n$ coefficients are expected to provide information on the initial state and the transport properties of the produced medium. The dominant coefficient is the second Fourier coefficient $v_2$, referred to as the elliptic flow~\cite{Ollitrault:1992bk,Poskanzer:1998yz}, which provides information on the collective expansion of the medium at low \pt~\cite{Kolb:2003dz} and the path-length dependence of medium-induced parton energy loss at high $\pT$~\cite{Gyulassy:2000gk,Shuryak:2001me}.

Collisions of small systems such as proton--nucleus are studied in detail to characterise the cold nuclear matter (CNM) effects which are also present in heavy-ion collisions, providing critical information for understanding QGP properties. These CNM effects influence parton distribution functions~\cite{Eskola:2009uj}, induce Cronin-like effects~\cite{Kopeliovich:2002yh}, and cause energy loss~\cite{Kang:2014hha}. As in heavy-ion collisions, a significant $v_{2}$ was observed in p--Pb collisions for soft as well as hard probes, such as open heavy-flavour hadrons~\cite{Acharya:2018dxy,Sirunyan:2020obi}, quarkonia~\cite{ALICE:2017smo,Sirunyan:2018kiz}, and high-\pt charged hadrons~\cite{Sirunyan:2017igb,Aad:2019ajj}. A positive $v_2$ was also reported for charged particles~\cite{ATLAS:2015hzw,ATLAS:2016yzd,Aaboud:2017acw,CMS:2017kcs,Acharya:2019vdf} and muons from charm-hadron decays~\cite{ATLAS:2019xqc} in pp collisions, while the $v_2$ of muons from beauty-hadron decays was consistent with zero~\cite{ATLAS:2019xqc}. 
The observation of a non-zero $v_2$ in small collision systems raised the question whether small-size QGP droplets are formed in these conditions.  
However, the particle yields at high \pt in p--Pb collisions are found to be unmodified within uncertainties compared to the same measurements in pp collisions, scaled by the number of binary nucleon--nucleon collisions~\cite{ALICE:2014xsp,ALICE:2015umm,ALICE:2016faw,Aad:2016zif,Acharya:2019hao,Aad:2019ajj,ALICE:2019fhe}. Such an observation indicates that final-state effects are not significant in small collision systems. Therefore, alternative scenarios have been proposed, including color exchange in the final state~\cite{Dusling:2012cg}, initial-state effects due to gluon saturation~\cite{Dusling:2013oia,Dusling:2015gta,Zhang:2020ayy}, or anisotropic escape of partons from the surface of the interaction zone~\cite{He:2015hfa}.

This paper reports the first measurements of the \pt-differential $v_2$ of jet particles (particles associated with jets) at midrapidity in high-multi\-pli\-ci\-ty \pPb collisions and semicentral \PbPb collisions at $\snn = 5.02$ TeV with ALICE. The jet-particle $v_2$ measured down to lower \pt compared to the $v_2$ of reconstructed jets is of particular interest since it is clearly separated from particles from soft processes. These results are compared with inclusive charged-particle $v_2$ measurements in both \pPb and \PbPb collisions, as well as with previous results of the $v_2$ of reconstructed jets in \PbPb collisions at $\snn = 2.76$~TeV~\cite{ALICE:2015efi}. Comparisons with A MultiPhase Transport (AMPT) model predictions~\cite{Lin:2004en, Li:2018leh} are also discussed.

The article is organised as follows. Section~\ref{sec:det} presents the ALICE apparatus with an emphasis on the detectors used in the analysis and the data taking conditions. The analysis strategy and the estimation of systematic uncertainties are described in Section~\ref{sec:ana}. Section~\ref{sec:res} presents the results for jet particles and inclusive charged particles measured in \pPb and  \PbPb collisions as well as comparisons with AMPT model calculations in \pPb collisions. Concluding remarks are drawn in Section~\ref{sec:concl}.


\section{Experimental apparatus and data samples}\label{sec:det}

The analysis is performed on the \pPb and \PbPb data collected with the ALICE detector in 2016 and 2015, respectively. In \pPb collisions, the asymmetry of the proton and Pb beam energies results in a rapidity shift of the nucleon--nucleon centre-of-mass by 0.465 in the direction of the proton beam with respect to the laboratory frame. In the following, the pseudorapidity $\eta$ values correspond to the laboratory frame. A detailed description of the detector and its performance is given in Refs.~\cite{Aamodt:2008zz,Abelev:2014ffa}. The analysis is based on tracks reconstructed using the Time Projection Chamber (TPC)~\cite{Alme:2010ke} located inside a large solenoidal magnet with a 0.5 T field parallel to the LHC beam direction and covering $\vert \eta \vert < 0.9$. Information from the Inner Tracking System (ITS)~\cite{ALICE:2010tia} is used to improve the spatial and momentum resolution of the reconstructed tracks. The Silicon Pixel Detector (SPD)~\cite{CERN-LHCC-99-012,ALICE:2010tia}, comprising the two innermost layers of the ITS covering $\vert \eta \vert < 2.0$ and $\vert \eta \vert < 1.4$, is employed together with the TPC to determine the position of the primary interaction vertex. The Forward Multiplicity Detector (FMD)~\cite{Cortese:781854} consists of three sets of silicon strip sensors, covering $-3.5 < \eta < -1.7$ (FMD3) and $1.7 < \eta < 5$ (FMD1,2). The FMD is used in \pPb collisions for the event selection and to extract the $v_2$ coefficient via long-range three-particle correlations. The V0 detector, formed by two scintillator arrays covering $-3.7 < \eta < -1.7$ (V0C) and $2.8 < \eta < 5.1$ (V0A), is used for triggering, event characterisation and centrality determination~\cite{ALICE:2013axi}. Two sets of Zero Degree Calorimeters (ZDC)~\cite{Abelev:2014ffa}, located at $\pm 112.5$~m from the nominal interaction point along the beam line, are also used for the event selection. 

The analysis of the \pPb and \PbPb samples is based on events selected by a minimum bias (MB) trigger. The MB trigger is provided by the coincidence of signals in the two V0 scintillator arrays. Pile-up events are removed based on an event selection which uses the information from the V0 and SPD to tag events with multiple vertices.
The beam-induced background is reduced offline by exploiting the V0 and ZDC timing information. In addition, in \pPb collisions, the correlation between the multiplicity measured in the FMD and V0 is used to further remove contamination from beam-induced background and outliers in the FMD multiplicity distribution. Only events with a primary vertex along the beam axis, $z_{\rm vtx}$, within $\pm 10$~cm from the nominal interaction point are considered. About 526 million \pPb and 60 million \PbPb events passed the event selection criteria. 
In \PbPb collisions, the centrality classes are defined as percentiles of the \PbPb hadronic cross section, and determined using the total energy deposited in the V0 arrays~\cite{ALICE-PUBLIC-2018-011}. The \pPb data sample is divided into several multiplicity classes based on the energy deposited in the V0A scintillators, located in the Pb-going direction~\cite{ALICE-PUBLIC-2018-011}. In the \pPb analysis, only the high-multiplicity class 0--10\% and the low-multiplicity class 60--100\% are studied. The measurements in \PbPb collisions are presented in the 20--60\% centrality interval.

Reconstructed charged-particle tracks are selected by applying the standard conditions given in Refs.~\cite{ALICE:2018vuu, ALICE:2018rtz, SupMat}. They concern the number of space points and the quality of the track fit in the TPC, and the distance of closest approach to the primary vertex.  
Tracks are selected with $\pt > 0.5$~GeV/$c$ and $\vert \eta \vert < 0.8$ in both \pPb and \PbPb collisions. Hits in the FMD are measured in the $\eta$ regions limited to $-3.2 < \eta < -1.8$ and $1.8 < \eta < 4.8$.


\section{Data analysis}\label{sec:ana}

The procedure developed to calculate the jet-particle $v_2$ is detailed in Ref.~\cite{SupMat}. The jet-particle $v_{2}$ measurement consists of four main parts: i) construction of the two-particle (charged particles at midrapidity regarded as trigger and associated particles) correlation function, 
ii) extraction of the near-side jet peak (the signal) and background yields containing also the away-side jet yields, by fitting the two-particle correlation function, iii) calculation of the $v_{2}$ of particle pairs represented by the trigger particles using three-particle correlations in p--Pb collisions, for the first time, and the scalar product method with the three-subevent technique in Pb--Pb collisions, and iv) extraction of the $v_{2}$ of jet particles, in given trigger- and associated-particle \pt intervals, using a two-component fit function that takes into account the relative contribution from both jets and background to the particle-pair $v_2$.

\subsection{Extraction of the near-side jet and background contributions}

\begin{figure}[!thb]
\begin{center}
\includegraphics[width=.47\textwidth]{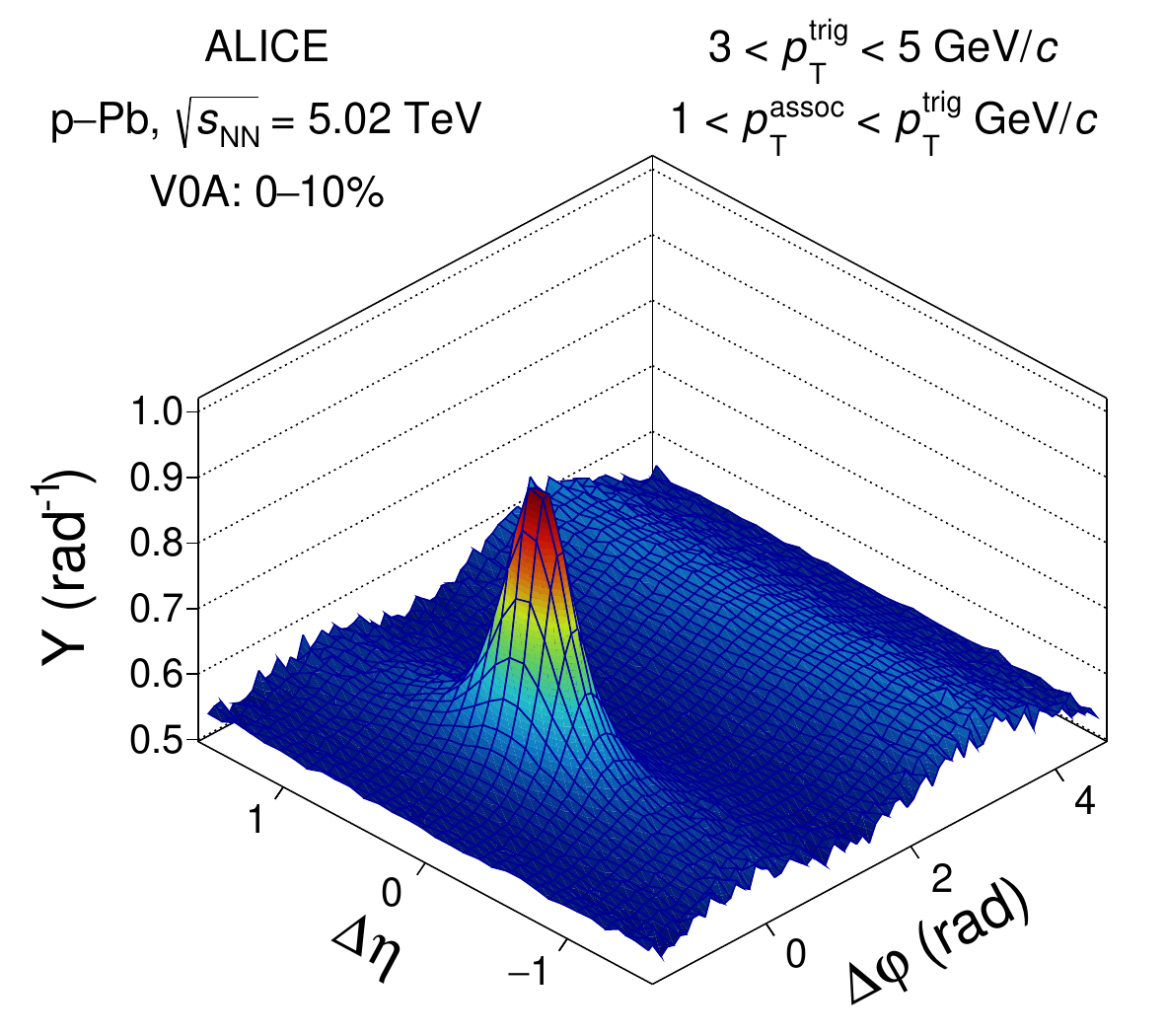}
\includegraphics[width=.47\textwidth]{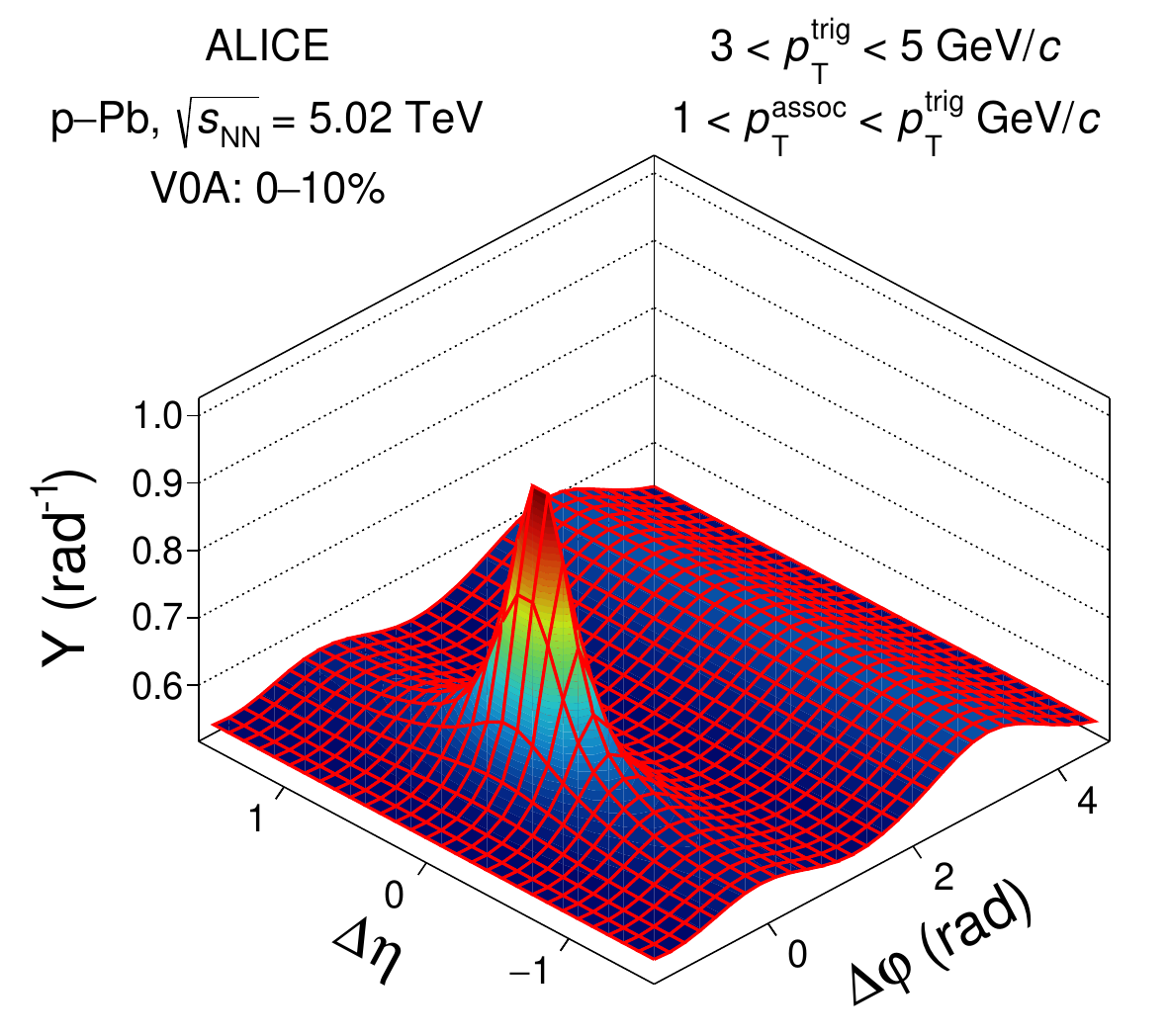}
\includegraphics[width=.47\textwidth]{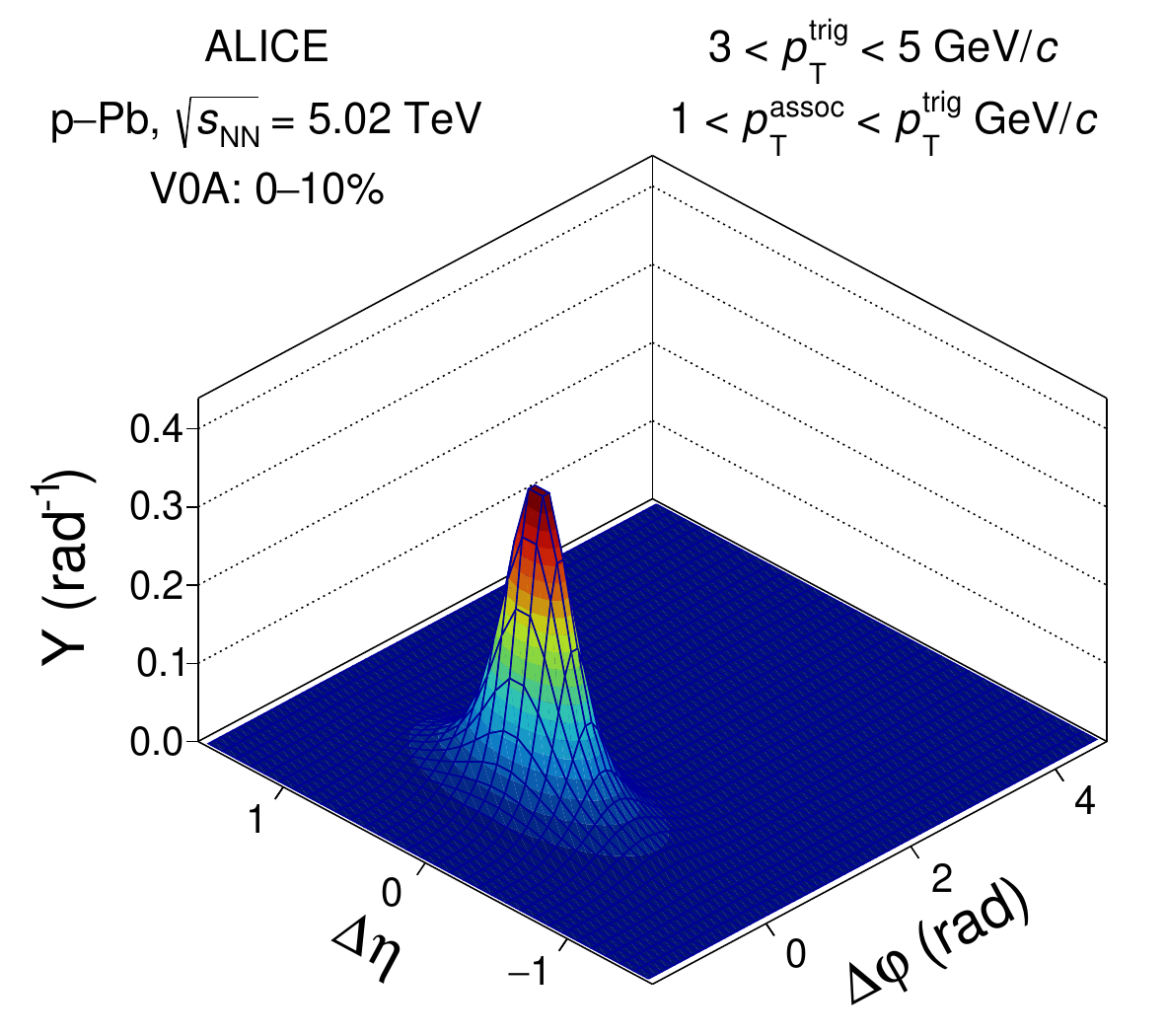}
\includegraphics[width=.47\textwidth]{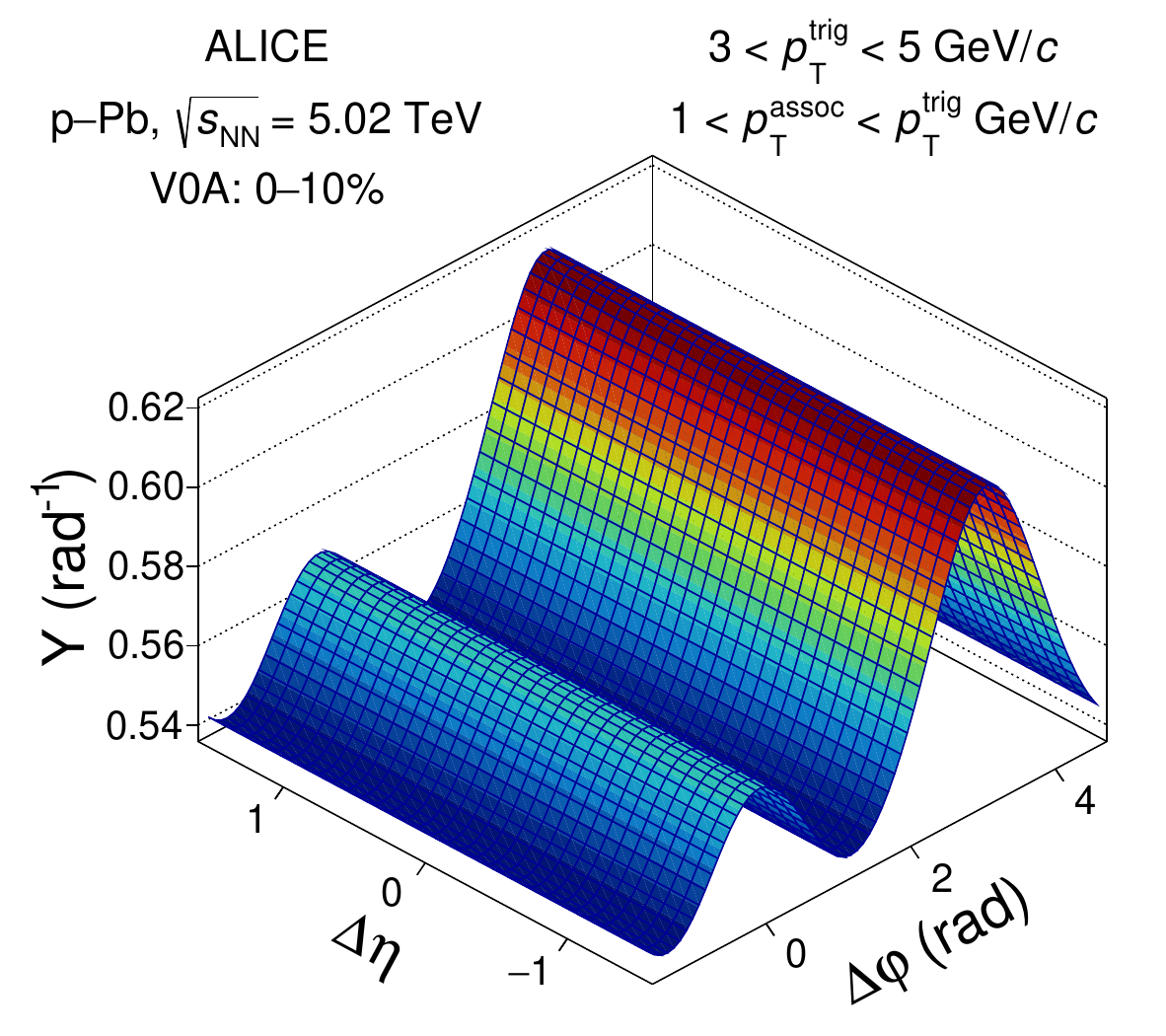}
\caption{Top: raw associated yield per trigger particle $Y$ (left) as a function of $\Delta \eta$ and $\Delta \varphi$ for pairs of charged particles measured in $\vert \eta \vert < 0.8$ with $3 < \pT^{\rm trig} < 5$~GeV/$c$ and $1 < \pT^{\rm assoc} < \pT^{\rm trig}$~GeV/$c$ in high-multiplicity (0--10\%) \pPb collisions at $\snn = 5.02$~TeV, and fit of the distribution (right). Bottom: extracted jet-particle (left) and background (right) yields.}
\label{Fig:2d-correlation}
\end{center}
\end{figure}
The two-dimensional correlation function is constructed as a function of the azimuthal angle difference ($\Delta \varphi$) and pseudorapidity difference ($\Delta \eta$) between trigger and associated particles (see Refs.~\cite{pubv2pPb,ALICE:2017smo} and references therein). Only pairs of particles with the same electric charge are considered to suppress correlations originating from resonance decays. The \pt of trigger particles ($\pt^{\rm trig}$) is taken to be larger than that of associated particles ($\pt^{\rm assoc}$) to avoid double counting of pairs. The correlation distribution is corrected for the limited two-particle acceptance and detector inhomogeneities by using the event-mixing technique~\cite{Abelev:2012ola}, and normalised to the total number of trigger particles. Therefore, the correlation function is expressed in terms of the associated yield per trigger particle $Y$ as
\begin{linenomath}
\begin{equation}
Y = \frac{1}{N_{\rm trig}} \frac{{\rm d}^2N_{\rm assoc}}{{\rm d}\Delta \eta {\rm d}\Delta \varphi} =  \frac{SE (\Delta \eta, \Delta \varphi)}{ME (\Delta \eta, \Delta \varphi)}, 
\label{eq:2pc}
\end{equation}
\end{linenomath}
where $N_{\rm trig}$ is the total number of trigger particles in a given event class. The signal distribution $SE(\Delta \eta, \Delta \varphi)$, given by $\displaystyle{\frac{1}{N_{\rm trig}} \frac{{\rm d}^2 N_{\rm same}}{{\rm d}\Delta \eta {\rm d} \Delta\varphi}}$, corresponds to the  associated yield per trigger particle for particle pairs from the same event, and the background distribution $ \displaystyle ME (\Delta \eta, \Delta \varphi) = \alpha \frac{{\rm d}^2 N_{\rm mix}}{{\rm d}\Delta \eta {\rm d} \Delta\varphi}$ is obtained by correlating trigger particles in an event with associated particles from other events of same event class. The $\alpha$ factor is introduced to normalise the background distribution to unity in the region of maximum pair acceptance. 
The same strategy is employed in \PbPb collisions. 

Figure~\ref{Fig:2d-correlation} (top-left) shows a typical example of the two-dimensional correlation distribution in high-multiplicity (0--10\%) \pPb collisions at $\snn = 5.02$~TeV for $3 < \pT^{\rm trig} < 5$~GeV/$c$ and $1 < \pT^{\rm assoc} < \pT^{\rm trig}$~GeV/$c$. A near-side jet structure is clearly observed at $(\Delta \varphi \sim 0, \Delta \eta \sim 0)$ on top of the background. The distribution is fitted with a double Gaussian and a sum of Fourier harmonics up to the fifth order~\cite{Abelev:2014mva} (Fig.~\ref{Fig:2d-correlation}, top-right). The former is used to extract the near-side jet-peak yield (Fig.~\ref{Fig:2d-correlation}, bottom-left), and the latter serves to obtain the background yield (Fig.~\ref{Fig:2d-correlation}, bottow-right).

\begin{figure}[!thb]
\begin{center}
\includegraphics[width=.47\textwidth]{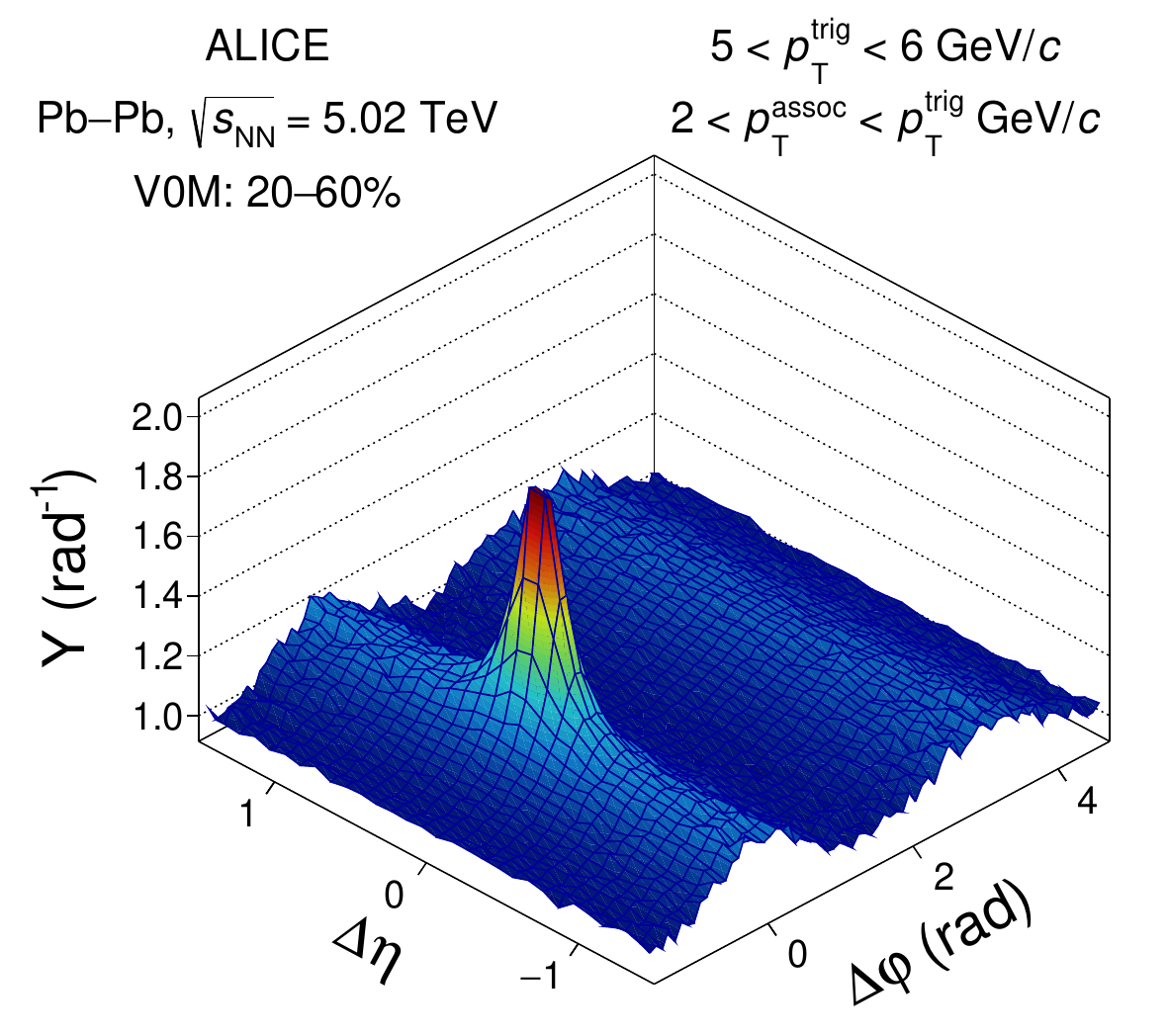}
\includegraphics[width=.47\textwidth]{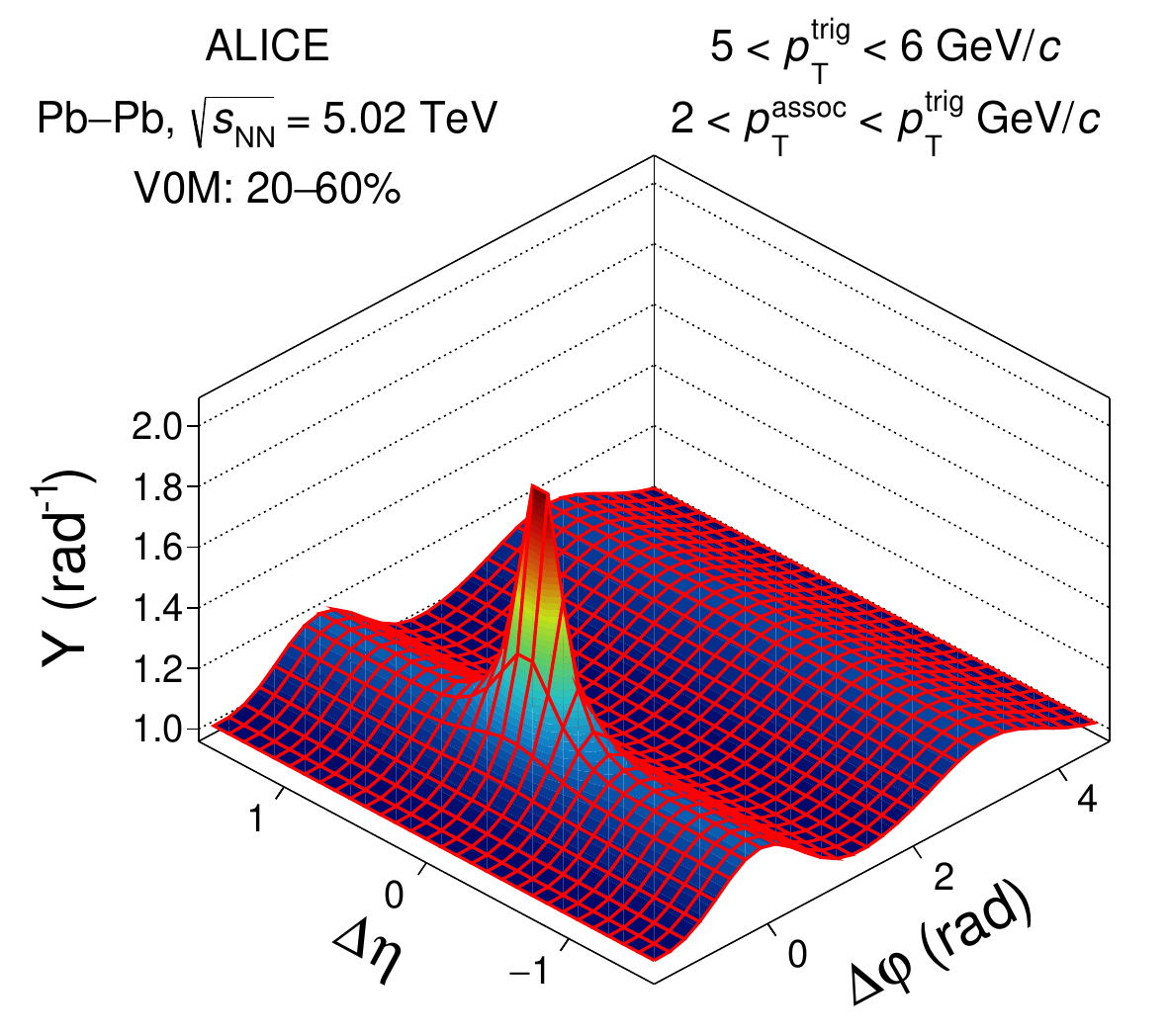}
\includegraphics[width=.47\textwidth]{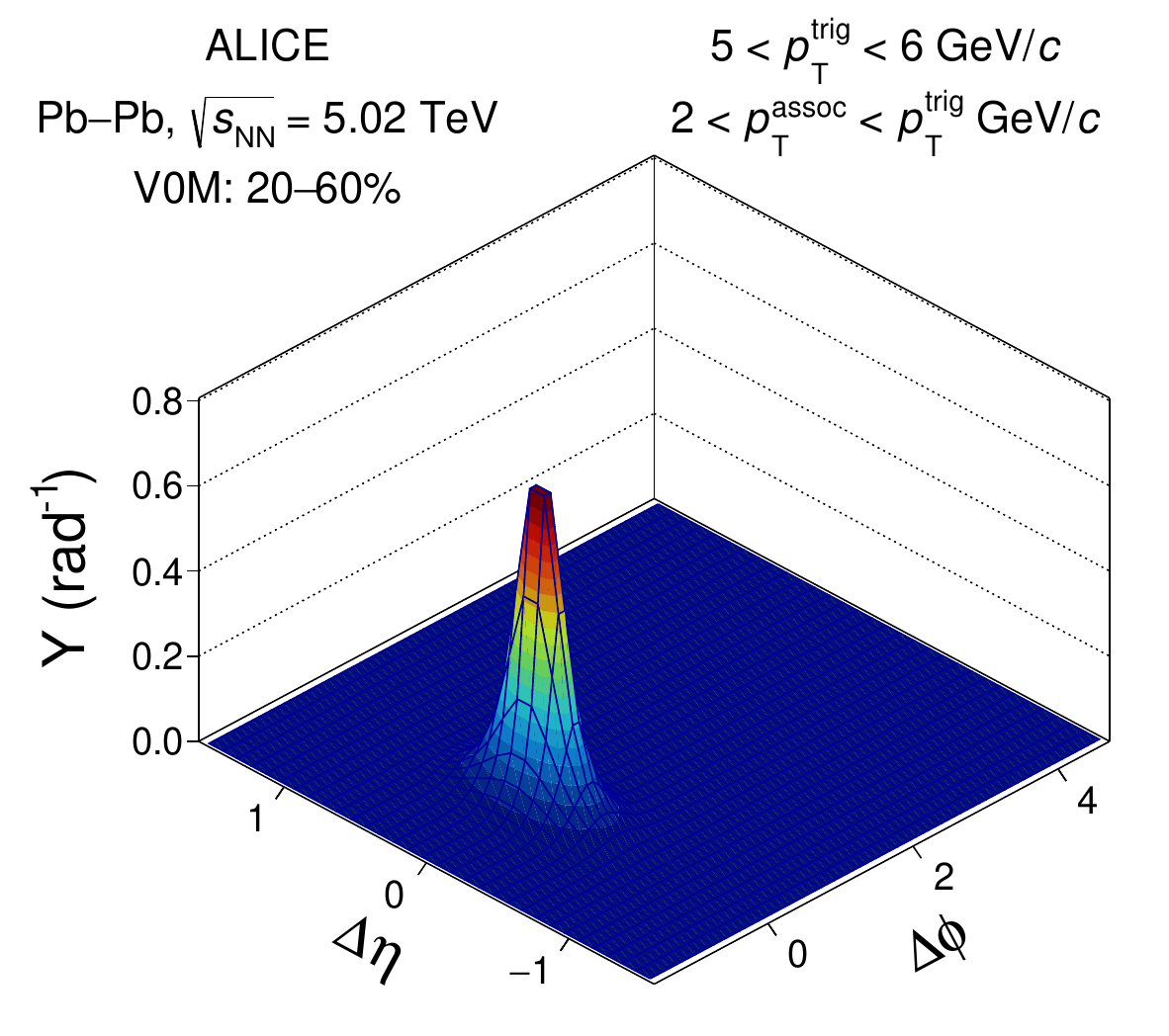}
\includegraphics[width=.47\textwidth]{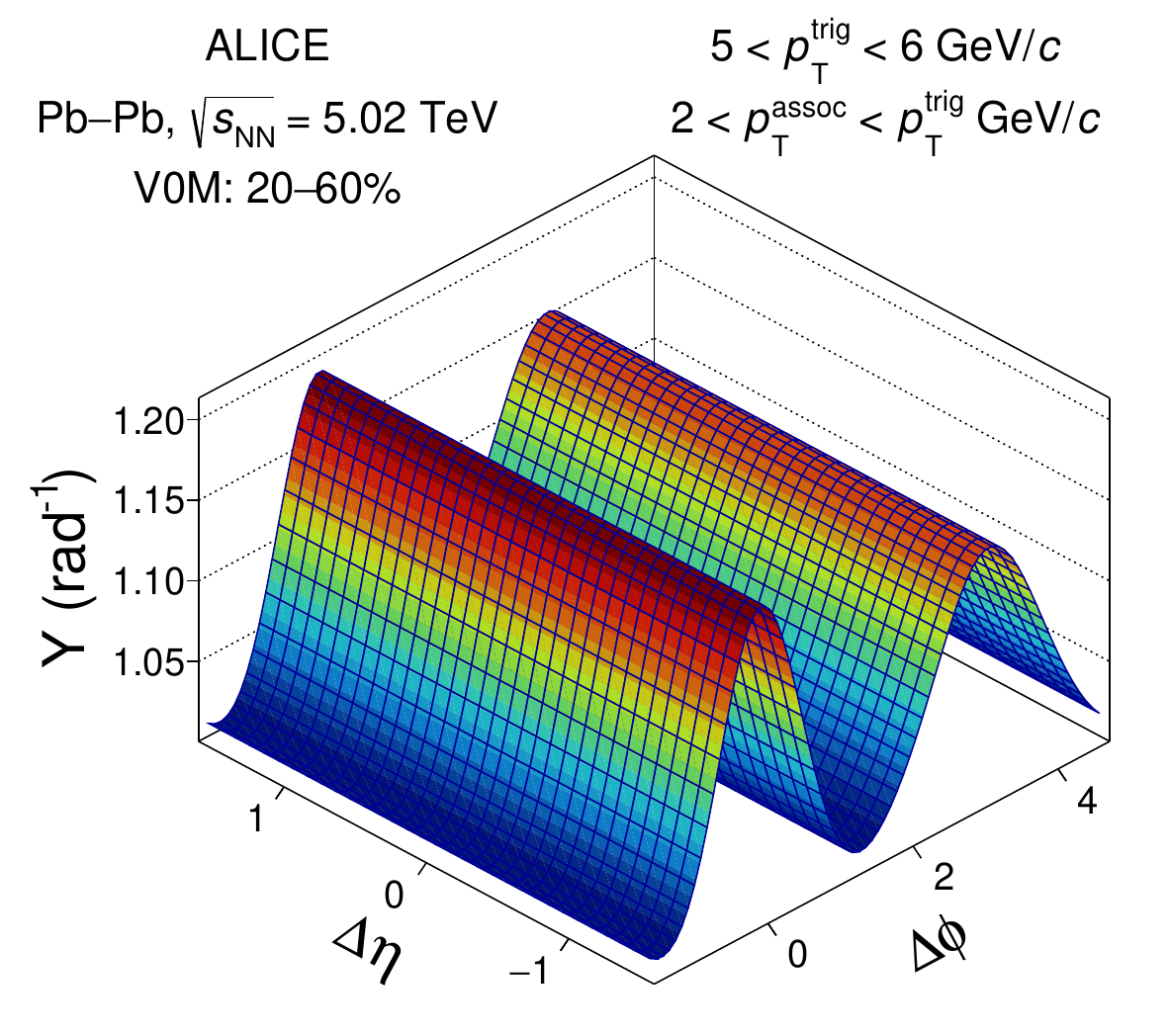}
\caption{Top: raw associated yield per trigger particle $Y$ (left) as a function of $\Delta \eta$ and $\Delta \varphi$ for pairs of charged particles measured in $\vert \eta \vert < 0.8$ with $5 < \pT^{\rm trig} < 6$~GeV/$c$ and $2 < \pT^{\rm assoc} < \pT^{\rm trig}$~GeV/$c$ in high-multiplicity (0--10\%) \PbPb collisions at $\snn = 5.02$~TeV, and fit of the distribution (right). Bottom: extracted jet-particle (left) and background (right) yields.}
\label{Fig:2d-correlation-PbPb}
\end{center}
\end{figure}

Similar typical distributions are also depicted in Fig.~\ref{Fig:2d-correlation-PbPb} for semicentral (20--60\%) \PbPb collisions at $\snn = 5.02$~TeV  for pairs of charged particles measured in $\vert \eta \vert < 0.8$ with $5 < \pT^{\rm trig} < 6$~GeV/$c$ and $2 < \pT^{\rm assoc} < \pT^{\rm trig}$~GeV/$c$ in semicentral (20--60\%) \PbPb collisions at $\snn = 5.02$~TeV. A near-side jet structure is also clearly observed at $(\Delta \varphi \sim 0, \Delta \eta \sim 0)$ on top of the background which is as expected, more important than in \pPb collisions. 

\subsection{Determination of the $v_2$ coefficient of particle pairs }

In the \pPb analysis, the $v_{2}$ of particle pairs, $v_2(\Delta \varphi, \Delta \eta$), is computed by using long-range three-particle correlations. The trigger particles in particle pairs for a given ($\Delta \varphi, \Delta \eta$) cell are correlated with particles selected in $1.8 < \eta < 4.8$ using FMD1,2 to construct the long-range correlation distribution as a function of their azimuthal angle difference ($\Delta \varphi^\prime$) and pseudorapidity difference ($\Delta \eta^\prime$). Nonflow contributions, such as dijets, are suppressed by subtracting the scaled ($\Delta \varphi^\prime, \Delta \eta^\prime$) correlation functions measured in low-multiplicity (60--100\%) collisions following the procedure described in Refs.~\cite{ATLAS:2016yzd, Adam:2015bka, pubv2pPb}, where the scaling factor is the ratio of the away-side jet yield in high-multiplicity collisions to that in low-multiplicity collisions. For each $\Delta \varphi^\prime$ interval, the $\Delta \eta^\prime$ distribution in the $-5.6 < \Delta \eta^\prime < -1.0$ range is integrated using a first-order polynomial fit to reduce the statistical fluctuations at the edges of $\Delta \eta^\prime$~\cite{Adam:2015bka}. This $\Delta \varphi^\prime$ distribution is fitted with a Fourier series parameterised with the first three harmonics as
\begin{linenomath}
  \begin{equation}
\frac{{\rm d}N}{{\rm d} \Delta\varphi^\prime} = a_0  (\Delta \varphi, \Delta \eta) + 2 \sum_{n = 1}^{3} a_{n} (\Delta \varphi, \Delta \eta) {\rm cos}(n\Delta \varphi^\prime) \propto 1 + 2 \sum_{n = 1}^{3}
  V_{n\Delta} (\Delta \varphi, \Delta \eta) {\mathrm{cos}}(n\Delta \varphi^\prime).
  \label{eq:3pc}
\end{equation}
\end{linenomath}

\begin{figure}[!thb]
\begin{center}
\includegraphics[width=.47\textwidth]{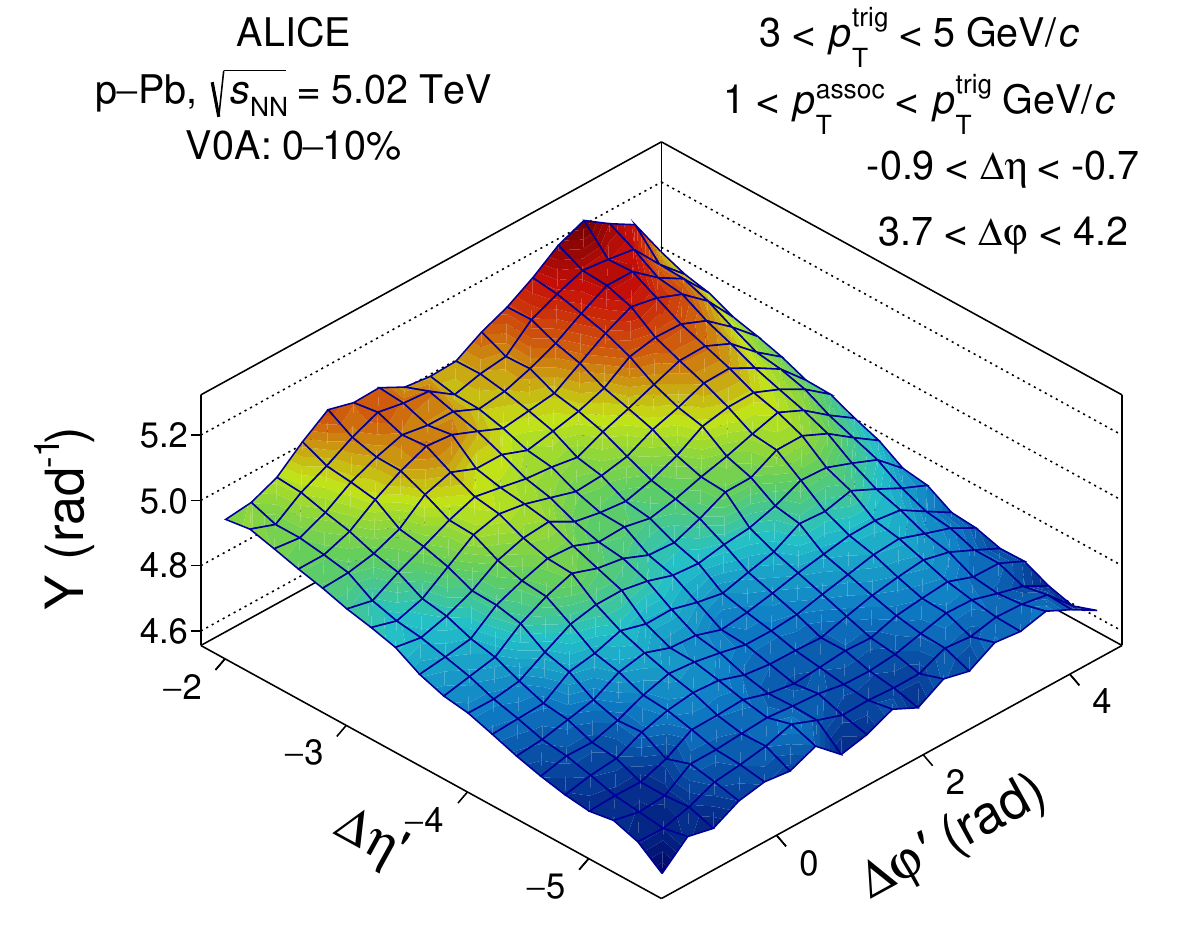}
\includegraphics[width=.47\textwidth]{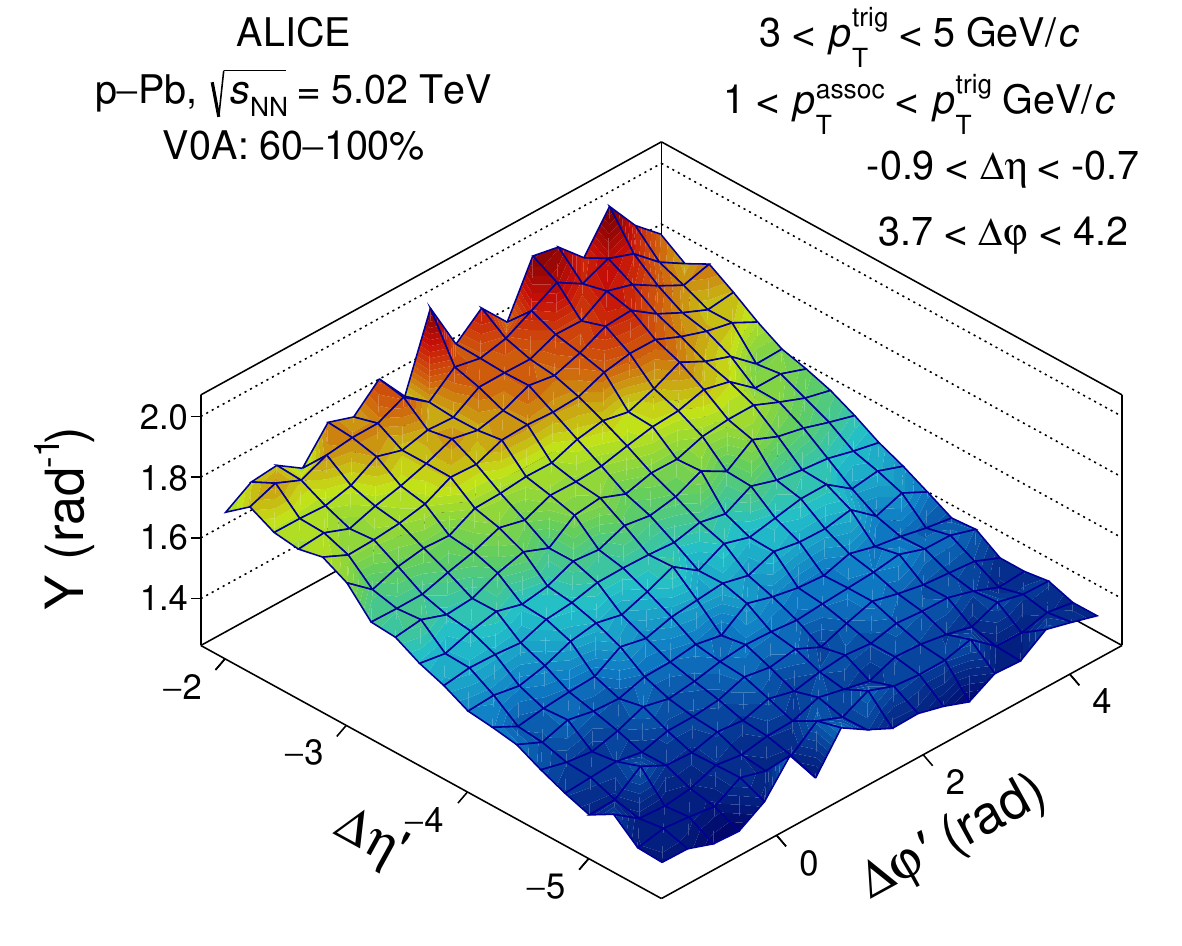}
\includegraphics[width=.47\textwidth]{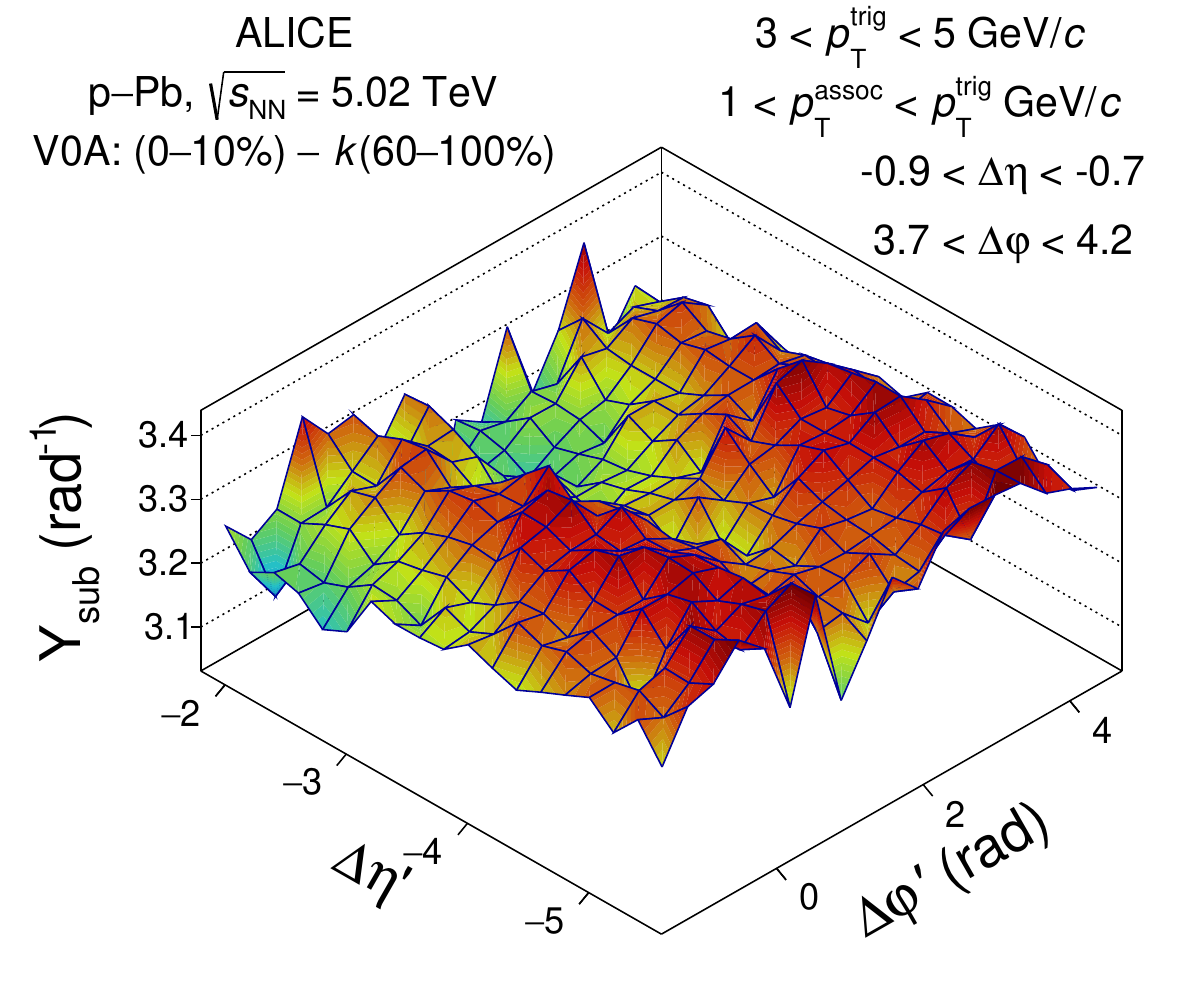}
\includegraphics[width=.47\textwidth]{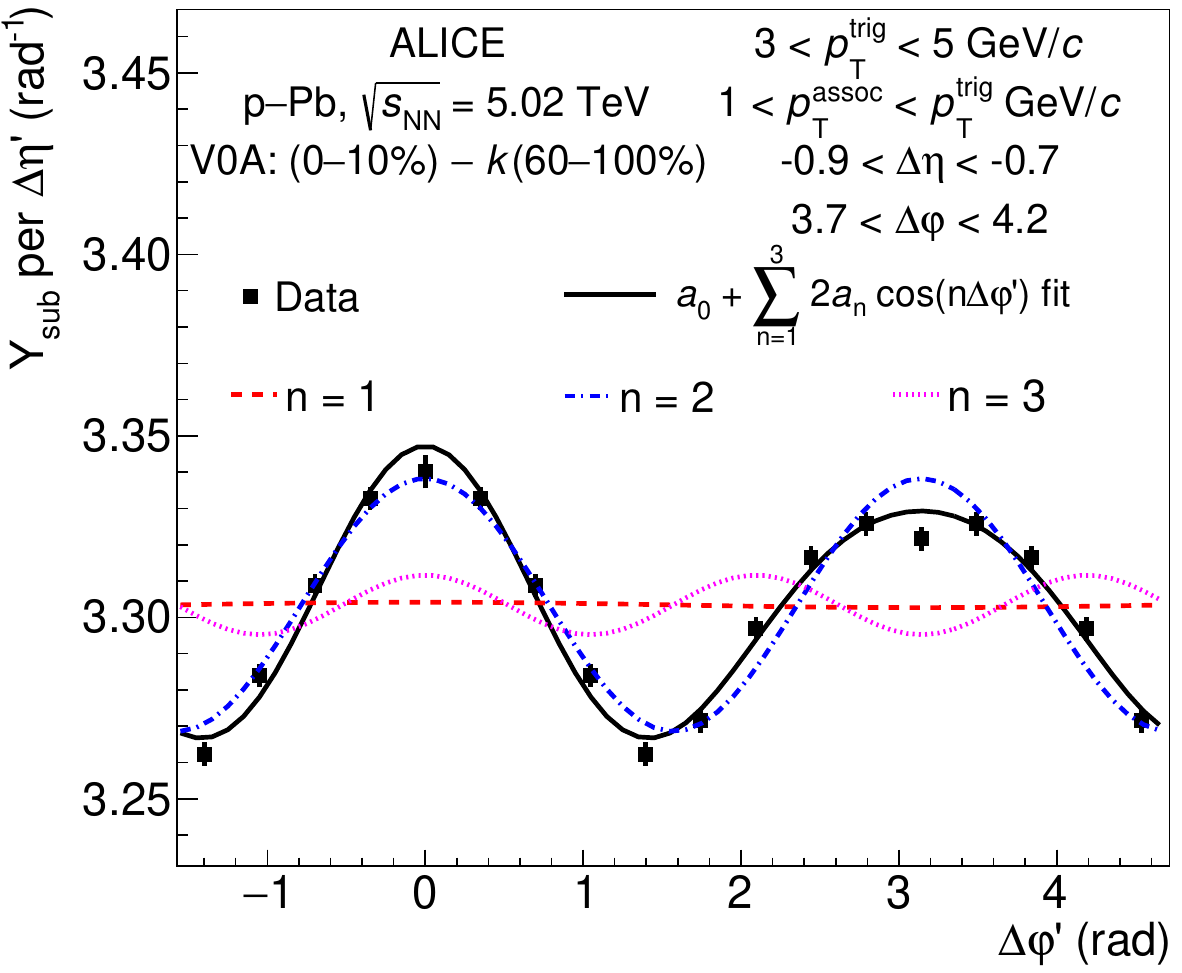}
\caption{Top: example of the associated yield per-trigger particle $Y$ in TPC--FMD1,2 correlations as a function of $\Delta \eta^\prime$ and $\Delta \varphi^\prime$ for $\vert \eta \vert < 0.8$ with $3 < \pT^{\rm trig} < 5$~GeV/$c$, $1 < \pT^{\rm assoc} < \pT^{\rm trig}$~GeV/$c$, $-0.9 < \Delta\eta < -0.7$ and $3.7 < \Delta\varphi < 4.2$ in high-multiplicity (left) and low-multiplicity (right) \pPb collisions at $\snn = 5.02$~TeV. Bottom: associated per-trigger yield after the subtraction of the scaled correlation distribution in low-multiplicity collisions(left)and fit of the distribution projected onto $\Delta\varphi^\prime$ with Eq.~(\ref{eq:3pc}).}
\label{Fig:3d-correlation}
\end{center}
\end{figure}

The $V_{2\Delta} (\Delta \varphi, \Delta \eta)$ second-order Fourier coefficient is extracted from the fit parameters and is further expressed relative to the baseline. The latter is estimated from the integral in $\Delta \varphi^\prime$ of the scaled correlation distribution in the low-multiplicity class around the minimum. The procedure is repeated for each \pt interval of trigger and associated charged particles.
Figure~\ref{Fig:3d-correlation} shows an example of the per-trigger associated yield as a function of $\Delta \eta^\prime$ and $\Delta \varphi^\prime$ for $3 < \pT^{\rm trig} < 5$~GeV/$c$, $1 < \pT^{\rm assoc} < \pT^{\rm trig}$~GeV/$c$, $-0.9 < \Delta \eta < -0.7$ and $3.7 < \Delta \varphi < 4.2$ in high-multiplicity (top-left) and low-multiplicity (top-right) \pPb collisions. The per-trigger associated yield in high-multiplicity \pPb collisions after the subtraction of the scaled correlation distribution in the low-multiplicity class and the fit of the corresponding distribution with Eq.~(\ref{eq:3pc}) are displayed in the bottom-left and bottom-right panels, respectively.

Assuming that $V_{2\Delta} (\Delta \varphi, \Delta \eta)$ can be factorised as the product of single-particle $v_{2}$ coefficients, the $v_2$ of particle pairs represented by trigger particles ($v_2(\Delta \varphi, \Delta \eta)$) is expressed as the ratio between the $V_{2\Delta} (\Delta \varphi, \Delta \eta)$ and the $v_{2}$ of particles in FMD1,2 ($v_2^{\rm FMD1,2}$). The $v_2^{\rm FMD1,2}$ is obtained with the three-subevent technique~\cite{Luzum:2012da} by constructing long-range two-particle correlations between trigger and associated particles in TPC--FMD1,2, TPC--FMD3 and FMD1,2--FMD3~\cite{ALICE:2023gyf}. If the factorisation holds, the $v_2^{\rm FMD1,2}$ which is given by
\begin{linenomath}
  \begin{equation}
    v_2^{\rm FMD1,2} = \sqrt{\frac{V_{2\Delta}^{\rm FMD1,2-FMD3} V_{2\Delta}^{\rm TPC-FMD1,2}}{V_{2\Delta}^{\rm TPC-FMD3}}}, 
  \label{eq:v2FMD}
  \end{equation}
  \end {linenomath}
amounts to $0.028$ (with negligible uncertainties) in the 0--10\% high-multiplicity class \pPb collisions. The effect of the contamination of secondary particles is discussed in Section~\ref{sec:syst}.

In \PbPb collisions, the $v_2 (\Delta \varphi, \Delta \eta)$ coefficient is extracted from the scalar product method via the three-subevent technique~\cite{STAR:2002hbo,Voloshin:2008dg}. The method correlates particle pairs measured in the TPC with the second-order event flow vector ${\bf Q}_2^{\rm{V0A}}$ estimated from the azimuthal distribution of the energy deposited in the V0A. Therefore, the resulting 
$v_2 (\Delta \varphi, \Delta \eta)$ is defined as 
\begin{linenomath}
\begin{equation}
v_2 (\Delta \varphi, \Delta \eta) = \big\langle {\bf u}_2 (\Delta \varphi,\Delta \eta)\cdot{\bf Q}_2^{\rm{V0A*}} \big\rangle_{\rm{TPC-TPC}} \bigg\slash 
\sqrt{\frac{\langle {\bf Q}_2^{\rm{V0A*}} \cdot {\bf Q}_2^{\rm{V0C}} \rangle
\langle {\bf Q}_2^{\rm{V0A}} \cdot {\bf Q}_2^{\rm{SPD*}}\rangle}
{\langle {\bf Q}_2^{\rm{V0C}}\cdot  {\bf Q}_2^{\rm{SPD*}}\rangle}},
\label{eq:v2-SP}
\end{equation}
\end{linenomath}
where ${\bf u}_2 (\Delta \varphi, \Delta \eta)$ is the unit flow vector of each particle measured in the TPC. The second-order harmonic event flow vectors ${\bf Q}_2^{\rm{V0C}}$ and ${\bf Q}_2^{\rm{SPD}}$ measured in the V0C and SPD, respectively, are introduced to take into account the resolution of the event flow vector ${\bf Q}_2^{\rm{V0A}}$. The symbol * represents the complex conjugate and the bracket $\langle ... \rangle_{\rm{TPC-TPC}}$ denotes the average over charged-particle pairs in a given \pt interval for trigger and associated particles, and centrality range. 
The brackets in the denominator denote the average over all events in a centrality class containing the particle pair. A recentering procedure is applied to correct the event flow vectors for the non-uniform azimuthal acceptance effects of the corresponding detectors~\cite{Selyuzhenkov:2007zi}. The pseudorapidity gaps between the TPC and V0A, and the V0A, V0C, and SPD detectors suppress nonflow effects~\cite{ALICE:2018rtz} and eliminate autocorrelations~\cite{ALICE:2018yph}. 

In both \pPb and \PbPb collisions, the $v_2(\Delta \varphi, \Delta \eta)$ can be written~\cite{ALICE:2018yph,ALICE:2013xna} as the weighted sum of the $v_{2}$ of jet particles ($v_2^{\rm jet\ part}$) and background ($v_2^{\rm B}(\Delta \varphi)$), as
\begin{linenomath}
\begin{equation}
v_2(\Delta \varphi, \Delta \eta) = \frac{S(\Delta \varphi, \Delta \eta)}{S(\Delta \varphi, \Delta \eta)+B(\Delta \varphi, \Delta \eta)} v_2^{\rm jet\ part} + \frac{B(\Delta \varphi, \Delta \eta)}{S(\Delta \varphi, \Delta \eta)+B(\Delta \varphi, \Delta \eta)} v_2^ {\rm B}(\Delta \varphi),
\label{eq:v2jet}
\end{equation}
\end{linenomath}
where the jet-particle $S(\Delta \varphi, \Delta \eta)$ and background $B(\Delta \varphi, \Delta \eta)$ yields are extracted from the two-particle correlation functions constructed in the TPC. The $v_2^{\rm jet\ part}$ coefficient is obtained by parametrising $v_2^{\rm B}(\Delta \varphi)$ with a Fourier series up to the fifth order and fitting Eq.~(\ref{eq:v2jet}) to the measured $v_2(\Delta \varphi, \Delta \eta)$ distributions in a given \pt interval for trigger and associated particles.

\begin{figure}[!thb]
\begin{center}
\includegraphics[width=.47\textwidth]{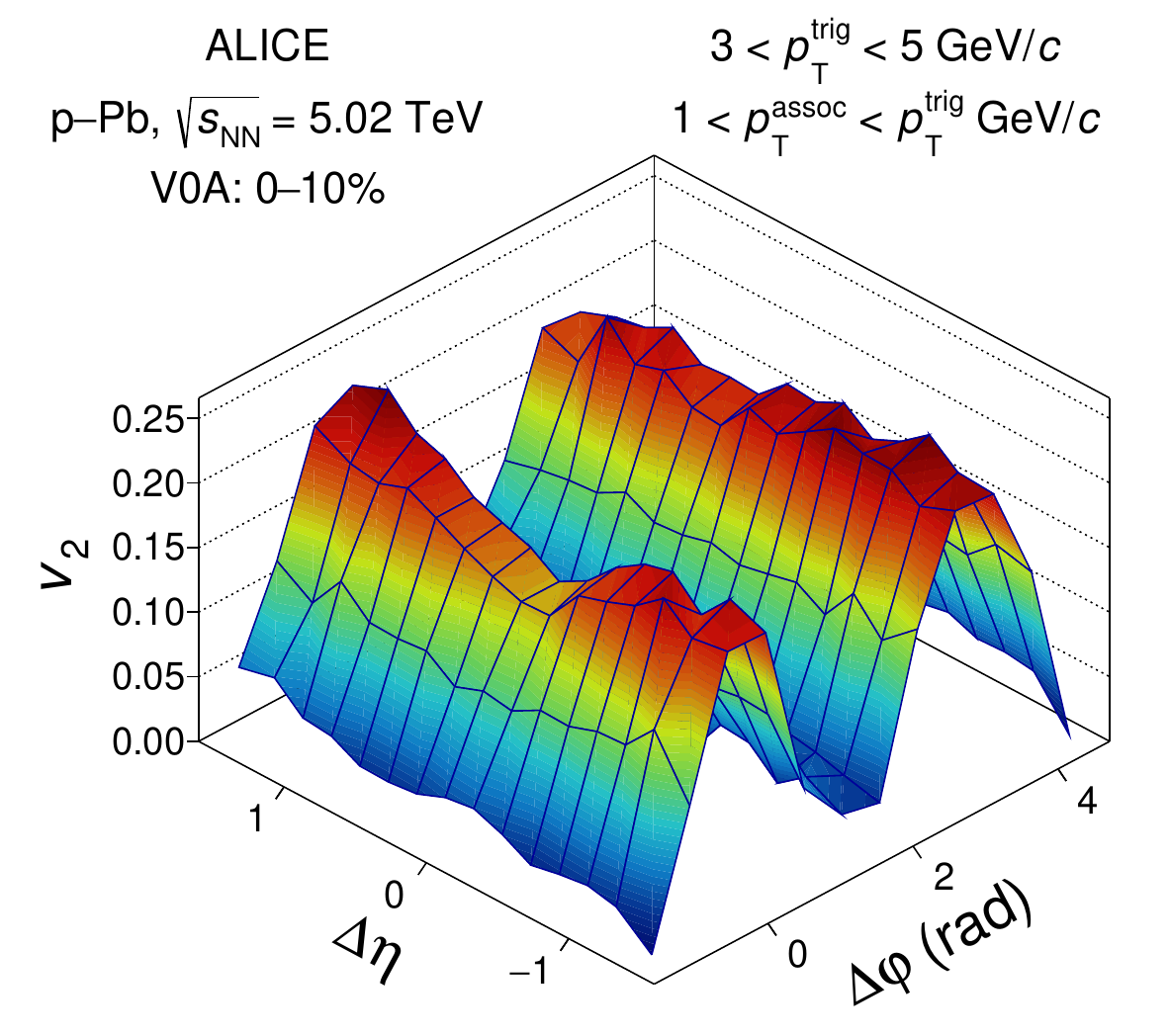}
\includegraphics[width=.47\textwidth]{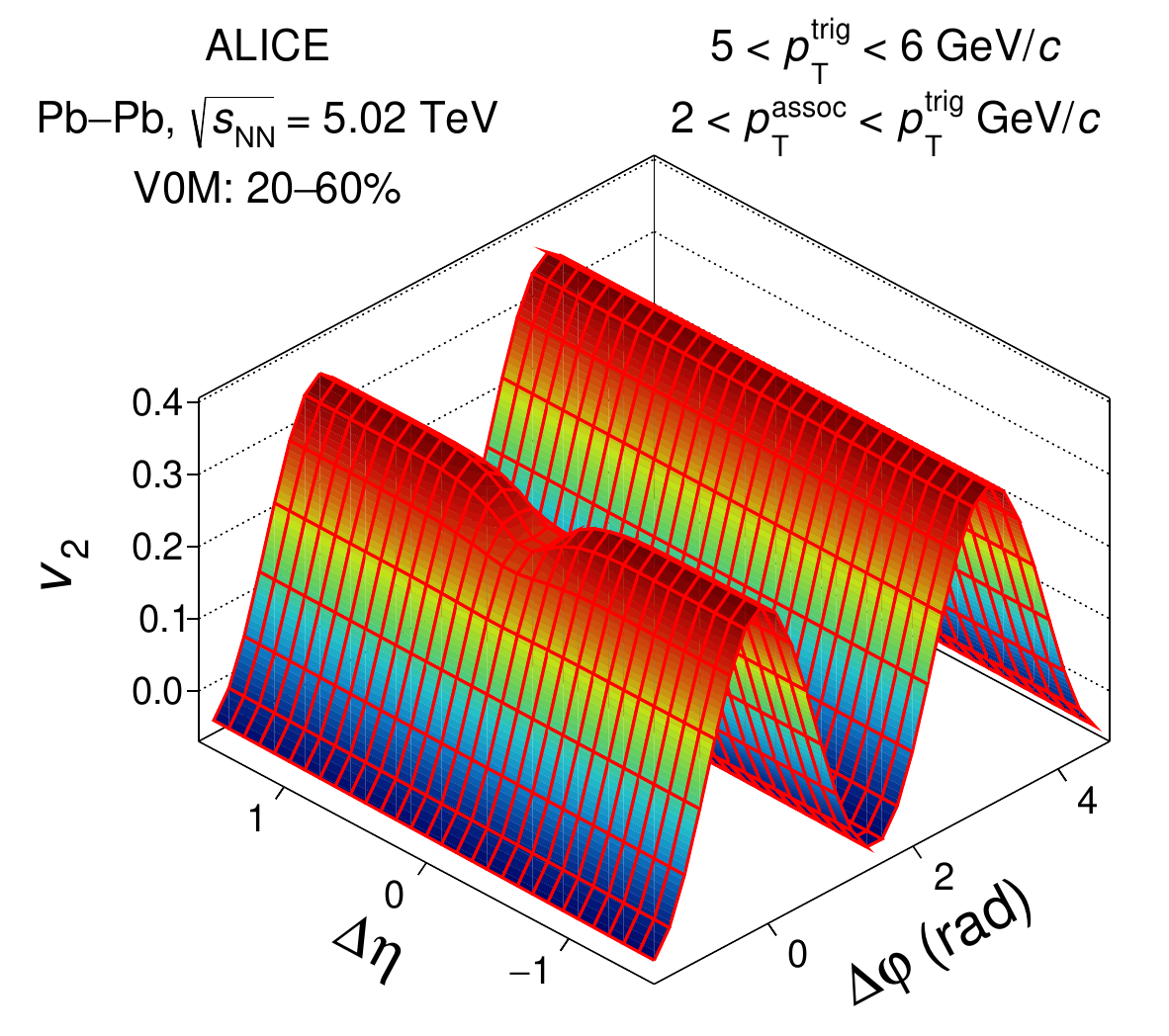}
\includegraphics[width=.47\textwidth]{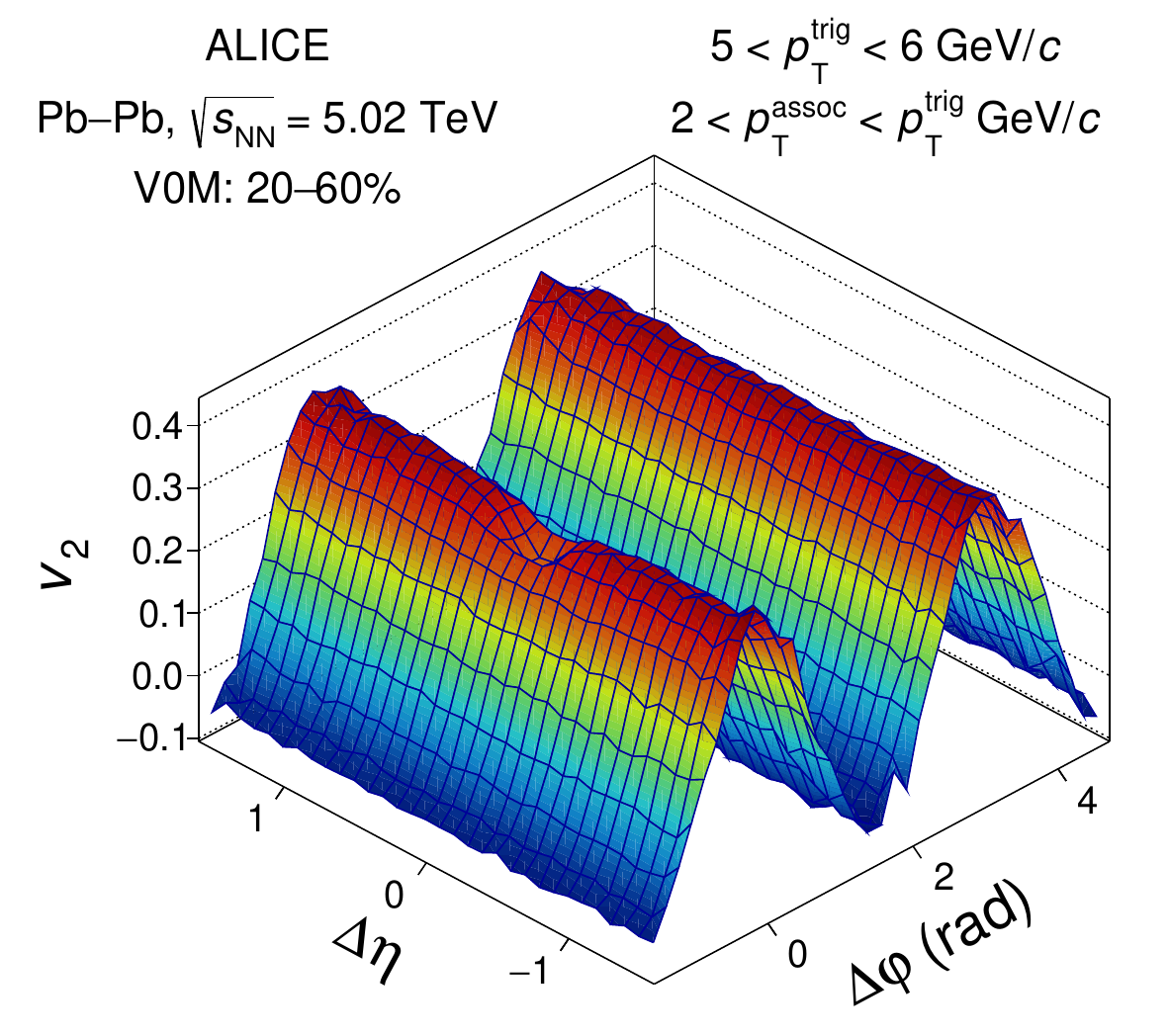}
\includegraphics[width=.47\textwidth]{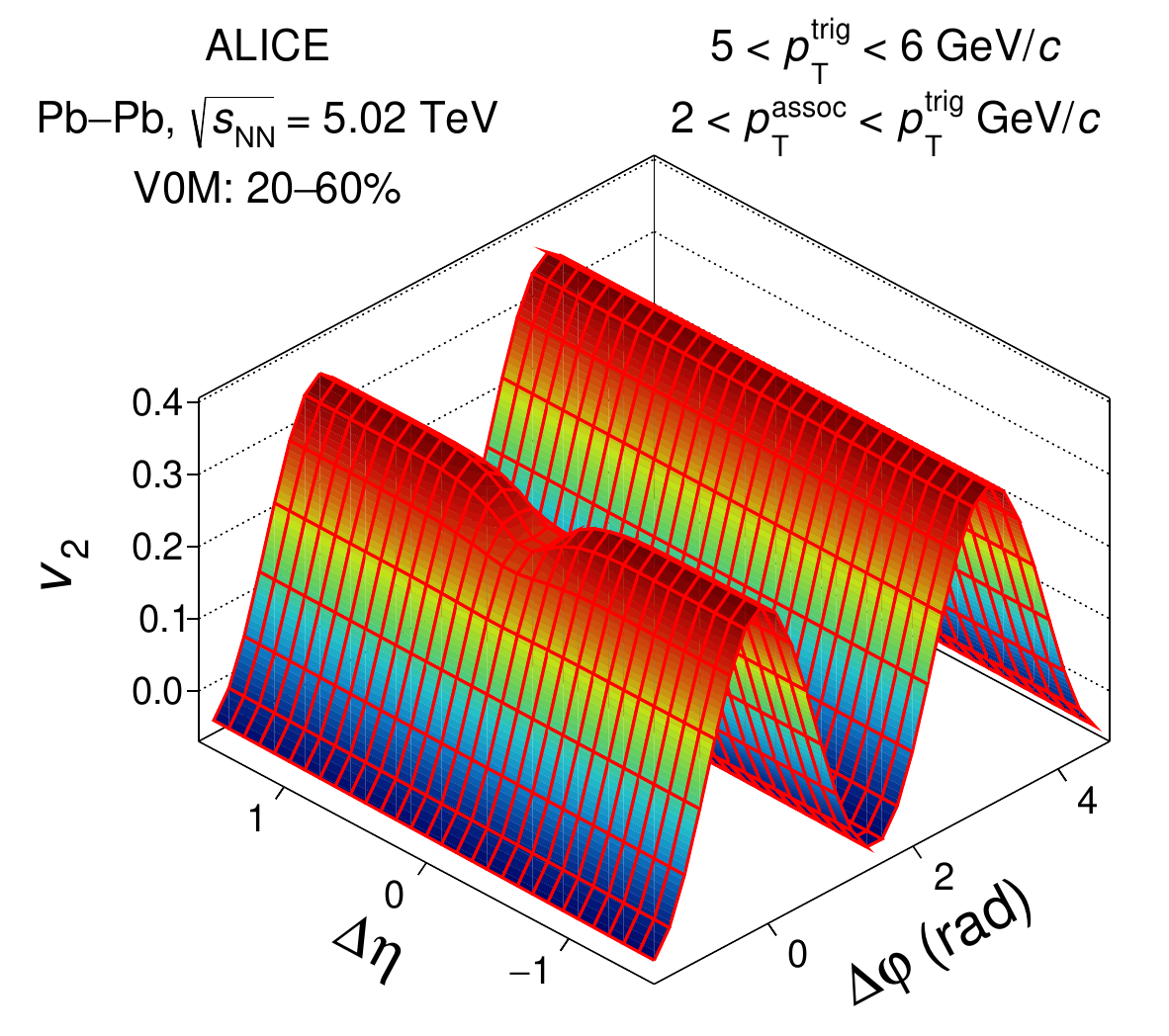}
\caption{Left: $v_2(\Delta \varphi, \Delta \eta)$ distributions of charged particles measured at midrapidity in high-multiplicity (0--10\%) \pPb collisions at $\snn = 5.02$~TeV (top) and semicentral (20--60\%) \PbPb collisions at $\snn = 5.02$~TeV (bottom). Right: fits fo the two corresponding distributions in \pPb (top) and \PbPb (bottom) collisions. The $\pT^{\rm trig}$ and $\pT^{\rm assoc}$ intervals are mentioned in the legend.}
\label{Fig:v2pairs}
\end{center}
\end{figure}

An example of the measured $v_2 (\Delta\varphi, \Delta\eta)$ distribution of particle pairs measured in \pPb collisions at $\snn = 5.02$~TeV for $3 < \pT^{\rm trig} < 5$~GeV/$c$ and $1 < \pT^{\rm assoc} < \pT^{\rm trig}$~GeV/$c$ is depicted in Fig.~\ref{Fig:v2pairs} (top-left), where a different structure is clearly seen in the region around the near-side jet peak ($\Delta \varphi \sim 0, \Delta \eta \sim 0$) compared to the background-dominated region. This is a first indication of a different behaviour for $v_2$ of jet and background particles. A  similar structure is also visible in \PbPb collisions at $\snn = 5.02$~TeV for $5 < \pT^{\rm trig} < 6$~GeV/$c$ and $2 < \pT^{\rm assoc} < \pT^{\rm trig}$~GeV/$c$ (Fig.~\ref{Fig:v2pairs}, bottom-left). The corresponding fits with Eq.~(\ref{eq:v2jet}) are shown in Fig.~\ref{Fig:v2pairs} (right panels).

For comparison, the $v_2$ coefficient of inclusive charged particles, $v_2^{\rm ch}$ is also computed. In \pPb collisions, it is evaluated using the three-subevent technique by constructing long-range two-particle correlations, as done for the $v_2^{\rm FMD1,2}$ calculation, and is expressed as
\begin{linenomath}
  \begin{equation}
    v_2^{\rm ch} = \sqrt{\frac{V_{2\Delta}^{\rm TPC-FMD1,2} V_{2\Delta}^{\rm TPC-FMD3}}{V_{2\Delta}^{\rm FMD1,2-FMD3  }}}.
  \label{eq:v2ch}
  \end{equation}
\end {linenomath}

In \PbPb collisions, the $v_2^{\rm ch}$ is determined using the scalar product method with the three-subevent technique (see Eq.~(\ref{eq:v2-SP})).

\subsection{Estimation of systematic uncertainties}\label{sec:syst}

Details of the separate contributions to the systematic uncertainties are given in Ref.~\cite{SupMat}. The values for both jet particles in a representative $\pt^{\rm assoc}$ interval and inclusive charged particles in \pPb and \PbPb collisions are summarised in Tables~\ref{Tab:pPb} and~\ref{Tab:PbPb}, respectively.

\begin{table}[!hbt]
  \caption{\label{Tab:pPb} Systematic uncertainties on the jet-particle $v_2$ for $p_{\rm T}^{\rm assoc} > 0.5$~GeV/$c$  and the inclusive charged-particle $v_2$ in high-multiplicity (0--10\%) \pPb collisions at $\snn = 5.02$~TeV. The systematic uncertainties vary within the indicated intervals depending on the \pt of trigger particles.}
\centering
\begin{tabular}{l|l|l}
Source & Jet particles & Charged particles \\
&$p_{\rm T}^{\rm assoc} $ > 0.5 GeV/$c$  & \\ \hline
Vertex selection & 0.6--14.6\% & 0.02--1.30\% \\
FMD-V0 correlation & 0.1--7.9\% & 0.1--0.2\% \\
Track selection & 0.3--3.6\% & 0.0--2.2\% \\
Secondaries in FMD  & 4\%& 4\% \\
Residual nonflow & 0.1--5.4\% & 0.1--13.2\% \\
Remaining ridge & 6.5--29.0\%& 0.7--20.7\% \\
$v_2$ calculation & 4.5--11.7\% & 0.7--1.5\% \\ \hline
Total & 11.2--34.3\% & 4.4--25.3\% \\
\end{tabular}
\end{table}

The jet-particle and inclusive charged-particle $v_2$ measurements in \pPb collisions are affected by the following systematic uncertainties. The variation of the range of $z_{\rm vtx}$ and a less stringent condition on the correlation between the multiplicity estimates obtained with the FMD and V0, give the systematic uncertainties related to the event selection. The uncertainty on the track reconstruction is estimated by modifying the track selection criteria. The bias due to the contribution of secondary particles produced in the FMD acceptance on the jet-particle $v_2$ is investigated in AMPT simulations~\cite{Lin:2004en,Li:2018leh,Lin:2021mdn}. In order to check for residual nonflow effects after the subtraction of the scaled low-multiplicity event class, the template fitting procedure~\cite{ATLAS:2016yzd} is tested. The difference between the two procedures is considered as a systematic uncertainty. A potential bias resulting from weak long-range correlations present in 60--100\% low-multiplicity events is studied by changing the interval from 60--100\% to 70--100\%. A systematic effect arises from the procedure employed for the $\Delta\varphi^\prime$. The $\Delta \varphi^{\prime}$ projection in Eq.~(\ref{eq:3pc}) is obtained from a constant fit instead of using a first-order polynomial fit along each $\Delta\eta^{\prime}$ interval. Finally, the baseline is also calculated in high-multiplicity collisions from the integral or from a second-order polynomial fit around the minimum at $\Delta\varphi^\prime \sim \pi/2$. The last two sources are the systematic uncertainty on the $v_2$ calculation. The aforementioned sources are added in quadrature in each \pt interval of trigger and associated particles to obtain a total systematic uncertainty on the jet-particle $v_2$ in the range 11.2--34.3\%. The total systematic uncertainty on the inclusive charged-particle which depends on the trigger-particle \pt, is 4.4--25.3\%.

\begin{table}[!hbt]
\caption{\label{Tab:PbPb} Systematic uncertainties on the jet-particle $v_2$ for $p_{\rm T}^{\rm assoc} > 2.0$~GeV/$c$ and the inclusive charged-particle $v_2$ in semicentral (20--60\%) \PbPb collisions at \snn = 5.02 TeV. The systematic uncertainties vary within the indicated intervals depending on the \pt of trigger particles.}
\centering
\begin{tabular}{l|l|l}
Source & Jet particles & Charged particles \\
& $p_{\rm T}^{\rm assoc} > 2$~GeV/$c$ &  \\ \hline
Vertex selection & 0.6--5.3\%& 0--5.3\% \\
Pile-up& 0.01--3.60\% & 0--2.7\% \\
Centrality & 0.3--1.4\% & 0--1.8\% \\
Flow vector& 0.3--1.4\% & 0--4.2\% \\
Track selection & 0.8--4.5\% & 0.01--1.80\% \\ \hline
Total & 1.6--7.3\% & 0.02--7.30\% \\
\end{tabular}
\end{table}

In \PbPb collisions, in addition to the systematic uncertainties arising from the variation of the $z_{\rm vtx}$ range and track selection criteria listed for \pPb collisions, a systematic uncertainty related to the centrality determination is estimated by using different centrality estimators. The systematic effect related to the pile-up event rejection is assessed via a dedicated analysis where pile-up events are not removed, only to estimate their importance. The event flow vector is computed by using V0C instead of V0A. All the systematic uncertainties are added in quadrature to obtain the total systematic uncertainty ranging from 1.6--10.1\% and 0.02--7.30\% for the jet-particle $v_2$ and inclusive charged-particle $v_2$, respectively.

\section{Results and model comparisons}\label{sec:res}

\begin{figure}[!hbt]
\begin{center}
\hspace*{-.2cm}\includegraphics[width=0.7\textwidth]{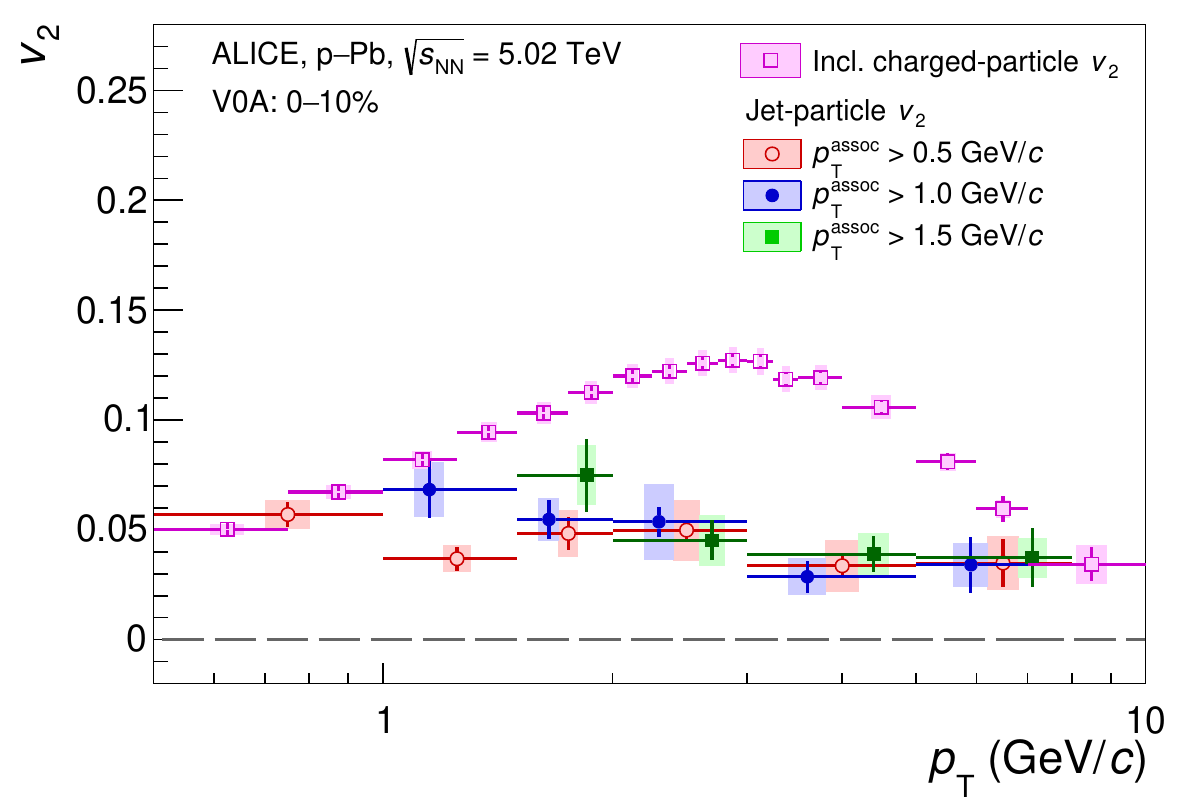}
\caption{Jet-particle $v_2$ as a function of the trigger-particle \pt for several $\pt^{\rm assoc}$ intervals in 0--10\% \pPb collisions at $\snn = 5.02$~TeV, compared with the inclusive charged-particle $v_2$. The values of the jet-particle $v_2$ are horizontally shifted around the centre of the bin for better visibility. The statistical uncertainties, shown as vertical bars, are determined using the sub-sample technique. The systematic uncertainties are represented as filled boxes. Horizontal bars indicate the bin width.}
\label{Fig:v2pPb}
\end{center}
\end{figure}

Figure~\ref{Fig:v2pPb} presents the $v_{2}$ of jet particles as a function of \pt at midrapidity ($\vert \eta \vert < 0.8$) for different $\pt^{\rm assoc}$ intervals, in high-multiplicity (0--10\%) \pPb collisions at $\snn = 5.02$ TeV. The \pt-differential inclusive charged-particle $v_2$ coefficient is also displayed. 
A positive jet-particle $v_2$ of $\sim$~0.04 is measured and it is independent of the \pt of trigger and associated particles within uncertainties. The significance is 2.6--6.8$\sigma$ for $\pt \lesssim 5$~GeV/$c$, depending on both the \pt of trigger and associated particles. In contrast, the $v_2$ of inclusive charged particles which mostly originates from the underlying event, is larger in magnitude and presents a clear dependence on \pt. This constitutes the first experimental evidence of different mechanisms in play for the $v_2$ of soft and hard probes at low \pt in \pPb collisions.

\begin{figure}[!tbh]
\begin{center}
\hspace*{-.2cm}\includegraphics[width=0.7\textwidth]{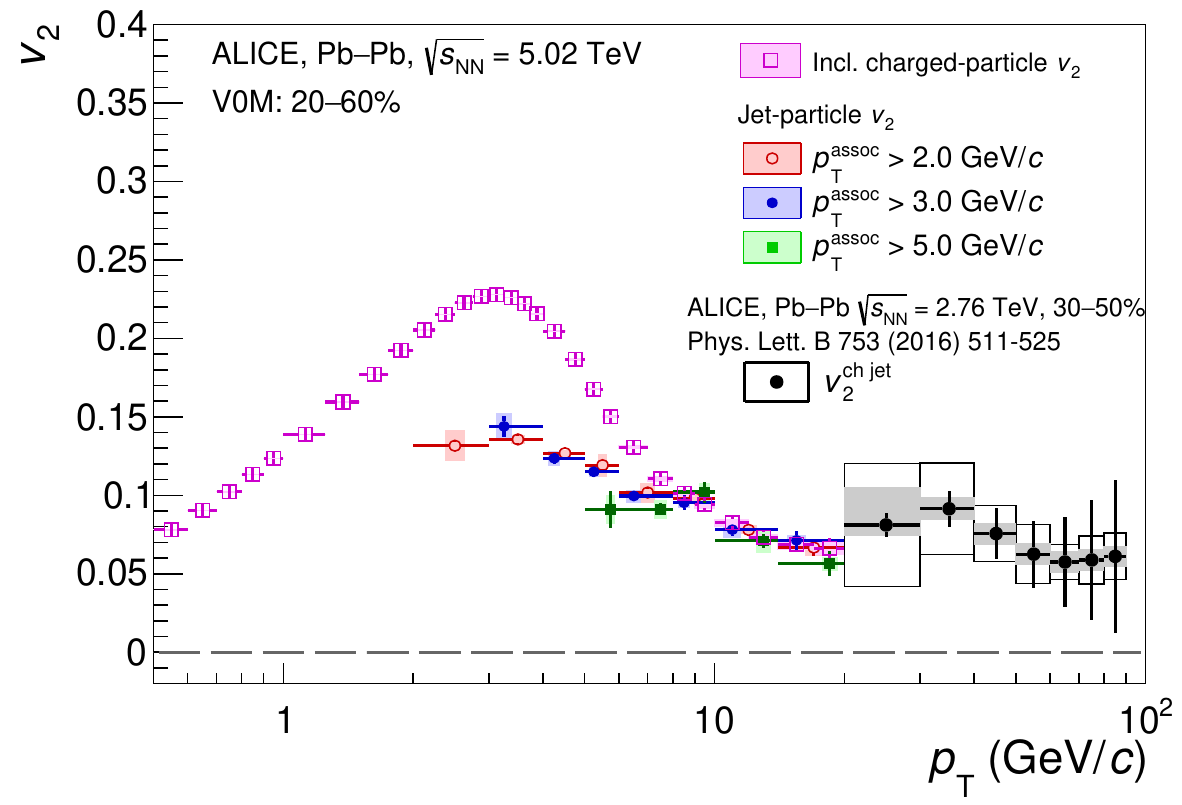}
\caption{Jet-particle $v_2$ as a function of the trigger-particle \pt for several $\pt^{\rm assoc}$ intervals in 20--60\% \PbPb collisions at $\snn = 5.02$ TeV, compared with the inclusive charged-particle $v_2$. The values of the jet-particle $v_2$ are horizontally shifted around the centre of the bin for better visibility. The statistical uncertainties, shown as vertical bars, are determined using the sub-sample technique. The systematic uncertainties are represented as filled boxes. Horizontal bars indicate the bin width. The published $v_2$ of reconstructed jets measured in 30--50\% \PbPb collisions is also shown~\cite{ALICE:2015efi}.}
\label{Fig:v2PbPb}
\end{center}
\end{figure}

The jet-particle $v_2$ and the inclusive charged-particle $v_2$ are also measured in semicentral (20--60\%) \PbPb collisions at $\snn = 5.02$ TeV, as shown in Fig.~\ref{Fig:v2PbPb}. The published $v_2$ of reconstructed jets measured in 30--50\% \PbPb collisions at $\snn = 2.76$ TeV~\cite{ALICE:2015efi} is also displayed, extending the presented measurements up to $\pt = 90$~GeV/$c$. A positive jet-particle $v_2$ is measured for the first time in the low \pt region, down to \pt = 2 GeV/$c$. The jet-particle $v_2$ does not exhibit any dependence on the associated \pt, while it  decreases with increasing \pt from \pt $\sim$ 3 GeV/$c$, and converges towards the $v_2$ of inclusive charged particles for $\pt \gtrsim 7$~GeV/$c$. In the high \pt region ($\pt \gtrsim 10$~GeV/$c$), the uniform behaviour of the inclusive charged-particle $v_2$ and jet $v_2$ as a function of \pt is attributed to the path-length dependent parton energy loss in \PbPb collisions~\cite{ALICE:2018vuu}. It can be noted that in this region, the jet-particle $v_2$ is consistent with the reconstructed jet $v_2$. In the interval $ 2 < \pt \lesssim 6$~GeV/$c$, the clear \pt dependence of the jet-particle $v_2$ in \PbPb collisions may be attributed to the interplay between hard partons and bulk particles.

\begin{figure}[!tbh]
\begin{center}
\hspace*{-.2cm}\includegraphics[width=0.7\textwidth]{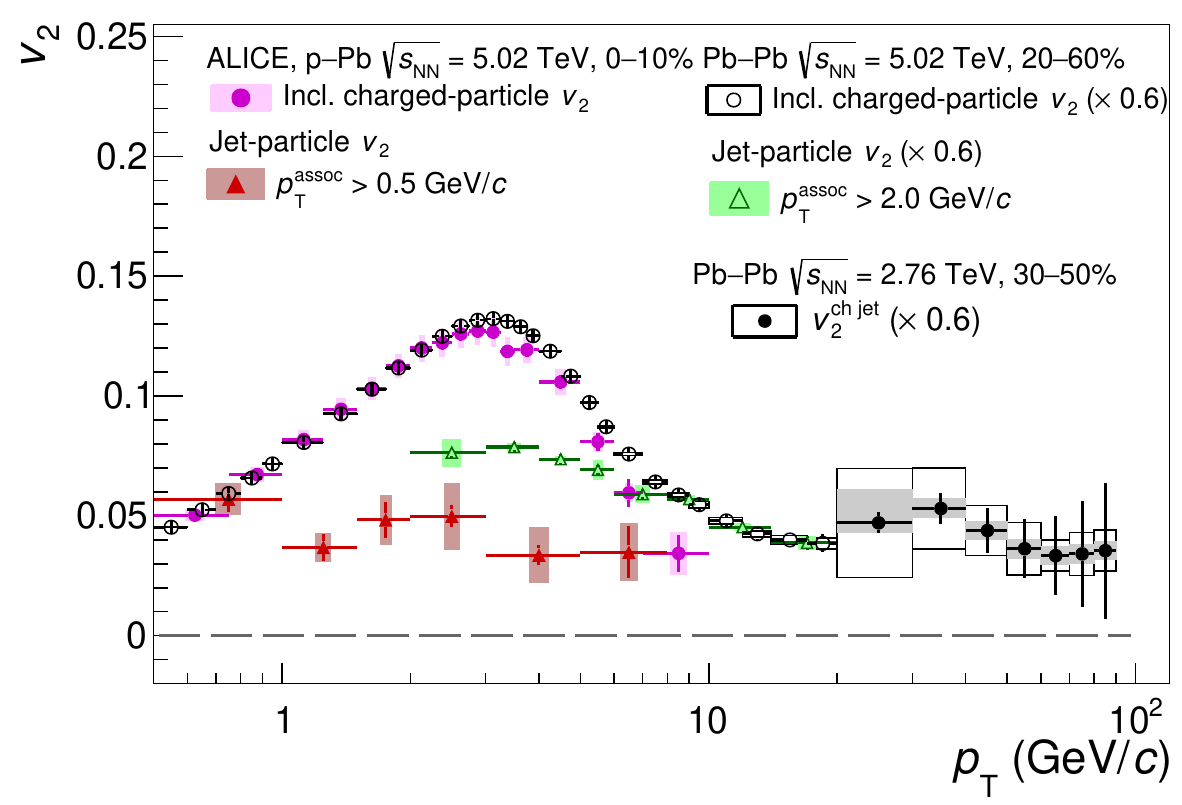}
\caption{Comparison of the \pt-differential jet-particle and inclusive charged-particle $v_2$ measured in high-multiplicity \pPb collisions at $\snn = 5.02$~TeV with the same measurements performed in semicentral \PbPb collisions. The published ALICE $v_2$ of reconstructed jets measured in 30--50\% \PbPb collisions is also shown~\cite{ALICE:2015efi}. The results obtained in \PbPb collisions are scaled by a factor 0.6. See the text for the details.}
\label{Fig:comp}
\end{center}
\end{figure}

The comparison of the $v_2$ results obtained in \pPb (0--10\%) and \PbPb (20--60\%) collisions for jet particles and inclusive charged particles is discussed in Fig.~\ref{Fig:comp}. Since no significant dependence on $\pt^{\rm assoc}$ is evidenced for jet particles (see Figs.~\ref{Fig:v2pPb} and~\ref{Fig:v2PbPb}),
the results are shown for the $\pt^{\rm assoc}$ interval which allows us to cover the largest \pt range. The centrality class is chosen such that the eccentricity, which quantifies the initial spatial anisotropies, is close to that in high-multiplicity \pPb collisions according to Glauber Monte Carlo simulations~\cite{Loizides:2017ack}, although the charged-particle multiplicities reached in the heavier system are larger. The published $v_2$ of reconstructed jets measured in 30--50\% \PbPb collisions at $\snn = 2.76$ TeV~\cite{ALICE:2015efi} are also included in the figure. A scaling factor of 0.6 is applied to the $v_2$ of inclusive charged particles, jet particles and reconstructed jets in \PbPb collisions to ensure that the low \pt ($\pt \lesssim 3$~GeV/$c$) inclusive charged-particle $v_2$ matches that in \pPb collisions. This factor may account for the slightly different spatial anisotropies in the two colliding systems and the larger multiplicity in the heaviest system~\cite{Aad:2019ajj}. After such scaling, the inclusive charged-particle $v_2$ in \pPb collisions is compatible with that measured in \PbPb collisions. A different behaviour is evidenced for jet particles.
In contrast to what is observed in \PbPb collisions, the jet-particle $v_2$ measured in \pPb collisions is found independent of \pt. This suggests that the collectivity from the initial state survives throughout all stages in the system evolution more easily in \pPb than \PbPb collisions~\cite{Wang:2022rdh}. This is also confirmed from the positive jet-particle $v_2$ measurement in \pPb collisions without any indication of a modification of the jet production yields within experimental uncertainties~\cite{Aad:2019ajj,ALICE:2016faw,ALICE:2015umm}.

\begin{figure}[!tbh]
\begin{center}
\hspace*{-.2cm}\includegraphics[width=0.7\textwidth]{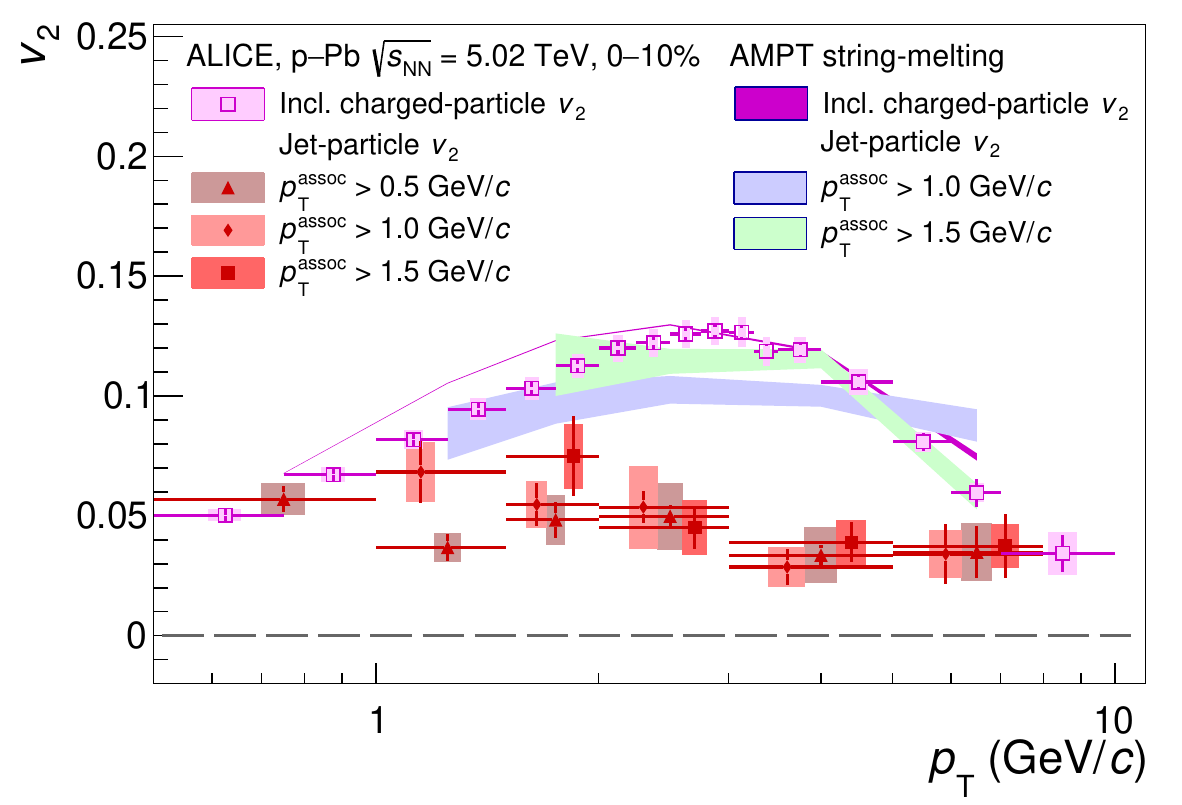}
\caption{Comparison of the \pt-differential jet-particle and inclusive charged-particle $v_2$ measured in high-multiplicity \pPb collisions at $\snn = 5.02$ TeV with AMPT calculations~\cite{Lin:2004en, Li:2018leh}.}
\label{Fig:AMPT}
\end{center}
\end{figure}

Figure~\ref{Fig:AMPT} presents a comparison of the jet-particle and inclusive charged-particle $v_2$ with the string-melting version of AMPT model (v2.26t9b)~\cite{Lin:2004en, Li:2018leh, Lin:2021mdn} in order to shed more light on the origin of the jet-particle $v_2$ measured in high-multiplicity \pPb collisions. The AMPT model includes four main processes: i) initial conditions obtained from the HIJING model~\cite{Wang:1991hta, Gyulassy:1994ew}, ii) parton scatterings based on the Zhang's parton cascade  (ZPC) model~\cite{Zhang:1997ej}, iii) hadronisation via coalescence and (iv) hadronic interactions described by a relativistic transport (ART) model~\cite{Li:1995pra}. A parton scattering cross section of 3 mb is used~\cite{Tang:2023wcd}. Its value is obtained by adjusting the Debye screening mass so that the model describes the \pt distribution and the $v_2$ coefficient of identified particles measured in \pPb collisions at \snn = 5.02 TeV by the ALICE Collaboration~\cite{ALICE:2016dei}. 
In this model, the $v_2$ is calculated following the same analysis procedure as with data and the event characterisation is done by mimicking the event class selection using the V0A detector at particle level, i.e. by counting charged particles in $2.8 < \eta < 5.1$. The AMPT calculations lead to a positive inclusive charged-particle and jet-particle $v_2$, indicating that parton interactions play an important role in the $v_2$ generation. The model provides a fair agreement with the measured inclusive charged-particle $v_2$, while it overestimates the measured jet-particle $v_2$, predicting a $v_2$ whose shape and magnitude are compatible with those of inclusive charged particles. This is possibly due to the fact that AMPT treats soft and hard components equally in the parton interaction stage, while the measurement indicates that hard partons interact with the underlying event differently from the bulk constituents themselves.

\section{Summary}\label{sec:concl}
In summary, the jet-particle $v_{2}$ measured in high-multiplicity (0--10\%) \pPb collisions at $\snn = 5.02$~TeV is assessed for the first time in the \pt range 0.5--8.0~GeV/$c$ by means of a novel multi-particle correlation technique. The jet-particle $v_2$ is also measured down to low \pt in semicentral \PbPb collisions at $\snn = 5.02$~TeV and is complementary to the previous jet $v_2$ results at higher \pt. Comparisons with the inclusive charged-particle $v_2$ measured in both \pPb and \PbPb collisions are discussed. A positive and \pt-independent $v_{2}$ signal is observed with a significance reaching 6.8$\sigma$ at low \pt in \pPb collisions. The $v_2$ magnitude is smaller than that measured in \PbPb collisions at intermediate \pt. In addition, a clear \pt dependence of the $v_2$ signal of jet particles is observed in \PbPb collisions. The comparison with AMPT predictions shows that parton interactions can generate a positive $v_{2}$ for jet particles in high-multiplicity \pPb collisions. 
These new results bring crucial information about the origin of the observed azimuthal anisotropies of jet particles in \pPb collisions and set key constraints on theoretical calculations.

\newenvironment{acknowledgement}{\relax}{\relax}
\begin{acknowledgement}
\section*{Acknowledgements}

The ALICE Collaboration would like to thank all its engineers and technicians for their invaluable contributions to the construction of the experiment and the CERN accelerator teams for the outstanding performance of the LHC complex.
The ALICE Collaboration gratefully acknowledges the resources and support provided by all Grid centres and the Worldwide LHC Computing Grid (WLCG) collaboration.
The ALICE Collaboration acknowledges the following funding agencies for their support in building and running the ALICE detector:
A. I. Alikhanyan National Science Laboratory (Yerevan Physics Institute) Foundation (ANSL), State Committee of Science and World Federation of Scientists (WFS), Armenia;
Austrian Academy of Sciences, Austrian Science Fund (FWF): [M 2467-N36] and Nationalstiftung f\"{u}r Forschung, Technologie und Entwicklung, Austria;
Ministry of Communications and High Technologies, National Nuclear Research Center, Azerbaijan;
Conselho Nacional de Desenvolvimento Cient\'{\i}fico e Tecnol\'{o}gico (CNPq), Financiadora de Estudos e Projetos (Finep), Funda\c{c}\~{a}o de Amparo \`{a} Pesquisa do Estado de S\~{a}o Paulo (FAPESP) and Universidade Federal do Rio Grande do Sul (UFRGS), Brazil;
Bulgarian Ministry of Education and Science, within the National Roadmap for Research Infrastructures 2020-2027 (object CERN), Bulgaria;
Ministry of Education of China (MOEC) , Ministry of Science \& Technology of China (MSTC) and National Natural Science Foundation of China (NSFC), China;
Ministry of Science and Education and Croatian Science Foundation, Croatia;
Centro de Aplicaciones Tecnol\'{o}gicas y Desarrollo Nuclear (CEADEN), Cubaenerg\'{\i}a, Cuba;
Ministry of Education, Youth and Sports of the Czech Republic, Czech Republic;
The Danish Council for Independent Research | Natural Sciences, the VILLUM FONDEN and Danish National Research Foundation (DNRF), Denmark;
Helsinki Institute of Physics (HIP), Finland;
Commissariat \`{a} l'Energie Atomique (CEA) and Institut National de Physique Nucl\'{e}aire et de Physique des Particules (IN2P3) and Centre National de la Recherche Scientifique (CNRS), France;
Bundesministerium f\"{u}r Bildung und Forschung (BMBF) and GSI Helmholtzzentrum f\"{u}r Schwerionenforschung GmbH, Germany;
General Secretariat for Research and Technology, Ministry of Education, Research and Religions, Greece;
National Research, Development and Innovation Office, Hungary;
Department of Atomic Energy Government of India (DAE), Department of Science and Technology, Government of India (DST), University Grants Commission, Government of India (UGC) and Council of Scientific and Industrial Research (CSIR), India;
National Research and Innovation Agency - BRIN, Indonesia;
Istituto Nazionale di Fisica Nucleare (INFN), Italy;
Japanese Ministry of Education, Culture, Sports, Science and Technology (MEXT) and Japan Society for the Promotion of Science (JSPS) KAKENHI, Japan;
Consejo Nacional de Ciencia (CONACYT) y Tecnolog\'{i}a, through Fondo de Cooperaci\'{o}n Internacional en Ciencia y Tecnolog\'{i}a (FONCICYT) and Direcci\'{o}n General de Asuntos del Personal Academico (DGAPA), Mexico;
Nederlandse Organisatie voor Wetenschappelijk Onderzoek (NWO), Netherlands;
The Research Council of Norway, Norway;
Commission on Science and Technology for Sustainable Development in the South (COMSATS), Pakistan;
Pontificia Universidad Cat\'{o}lica del Per\'{u}, Peru;
Ministry of Education and Science, National Science Centre and WUT ID-UB, Poland;
Korea Institute of Science and Technology Information and National Research Foundation of Korea (NRF), Republic of Korea;
Ministry of Education and Scientific Research, Institute of Atomic Physics, Ministry of Research and Innovation and Institute of Atomic Physics and Universitatea Nationala de Stiinta si Tehnologie Politehnica Bucuresti, Romania;
Ministry of Education, Science, Research and Sport of the Slovak Republic, Slovakia;
National Research Foundation of South Africa, South Africa;
Swedish Research Council (VR) and Knut \& Alice Wallenberg Foundation (KAW), Sweden;
European Organization for Nuclear Research, Switzerland;
Suranaree University of Technology (SUT), National Science and Technology Development Agency (NSTDA) and National Science, Research and Innovation Fund (NSRF via PMU-B B05F650021), Thailand;
Turkish Energy, Nuclear and Mineral Research Agency (TENMAK), Turkey;
National Academy of  Sciences of Ukraine, Ukraine;
Science and Technology Facilities Council (STFC), United Kingdom;
National Science Foundation of the United States of America (NSF) and United States Department of Energy, Office of Nuclear Physics (DOE NP), United States of America.
In addition, individual groups or members have received support from:
European Research Council, Strong 2020 - Horizon 2020, Marie Sk\l{}odowska Curie (grant nos. 950692, 824093, 896850), European Union;
ICSC - Centro Nazionale di Ricerca in High Performance Computing, Big Data and Quantum Computing, European Union - NextGenerationEU;
Academy of Finland (Center of Excellence in Quark Matter) (grant nos. 346327, 346328), Finland;
Programa de Apoyos para la Superaci\'{o}n del Personal Acad\'{e}mico, UNAM, Mexico.
\end{acknowledgement}

\bibliographystyle{utphys} 
\bibliography{AliJetv2pPb}

\newpage
\appendix

\section{The ALICE Collaboration}
\label{app:collab}
\begin{flushleft} 
\small

S.~Acharya\,\orcidlink{0000-0002-9213-5329}\,$^{\rm 126}$, 
D.~Adamov\'{a}\,\orcidlink{0000-0002-0504-7428}\,$^{\rm 86}$, 
A.~Adler$^{\rm 70}$, 
G.~Aglieri Rinella\,\orcidlink{0000-0002-9611-3696}\,$^{\rm 32}$, 
M.~Agnello\,\orcidlink{0000-0002-0760-5075}\,$^{\rm 29}$, 
N.~Agrawal\,\orcidlink{0000-0003-0348-9836}\,$^{\rm 51}$, 
Z.~Ahammed\,\orcidlink{0000-0001-5241-7412}\,$^{\rm 134}$, 
S.~Ahmad\,\orcidlink{0000-0003-0497-5705}\,$^{\rm 15}$, 
S.U.~Ahn\,\orcidlink{0000-0001-8847-489X}\,$^{\rm 71}$, 
I.~Ahuja\,\orcidlink{0000-0002-4417-1392}\,$^{\rm 37}$
% ...

\section*{Affiliation Notes}

$^{\rm I}$ Deceased\\
$^{\rm II}$ Also at: Max-Planck-Institut fur Physik, Munich, Germany\\
$^{\rm III}$ Also at: Italian National Agency for New Technologies, Energy and Sustainable Economic Development (ENEA), Bologna, Italy\\
$^{\rm IV}$ Also at: Dipartimento DET del Politecnico di Torino, Turin, Italy\\
$^{\rm V}$ Also at: Yildiz Technical University, Istanbul, T\"{u}rkiye\\
$^{\rm VI}$ Also at: Department of Applied Physics, Aligarh Muslim University, Aligarh, India\\
$^{\rm VII}$ Also at: Institute of Theoretical Physics, University of Wroclaw, Poland\\
$^{\rm VIII}$ Also at: An institution covered by a cooperation agreement with CERN\\

\section*{Collaboration Institutes}

$^{1}$ A.I. Alikhanyan National Science Laboratory (Yerevan Physics Institute) Foundation, Yerevan, Armenia\\
$^{2}$ AGH University of Krakow, Cracow, Poland\\
$^{3}$ Bogolyubov Institute for Theoretical Physics, National Academy of Sciences of Ukraine, Kiev, Ukraine\\
$^{4}$ Bose Institute, Department of Physics  and Centre for Astroparticle Physics and Space Science (CAPSS), Kolkata, India\\
$^{5}$ California Polytechnic State University, San Luis Obispo, California, United States\\
$^{6}$ Central China Normal University, Wuhan, China\\
$^{7}$ Centro de Aplicaciones Tecnol\'{o}gicas y Desarrollo Nuclear (CEADEN), Havana, Cuba\\
$^{8}$ Centro de Investigaci\'{o}n y de Estudios Avanzados (CINVESTAV), Mexico City and M\'{e}rida, Mexico\\
$^{9}$ Chicago State University, Chicago, Illinois, United States\\
$^{10}$ China Institute of Atomic Energy, Beijing, China\\
$^{11}$ Chungbuk National University, Cheongju, Republic of Korea\\
$^{12}$ Comenius University Bratislava, Faculty of Mathematics, Physics and Informatics, Bratislava, Slovak Republic\\
$^{13}$ COMSATS University Islamabad, Islamabad, Pakistan\\
$^{14}$ Creighton University, Omaha, Nebraska, United States\\
$^{15}$ Department of Physics, Aligarh Muslim University, Aligarh, India\\
$^{16}$ Department of Physics, Pusan National University, Pusan, Republic of Korea\\
$^{17}$ Department of Physics, Sejong University, Seoul, Republic of Korea\\
$^{18}$ Department of Physics, University of California, Berkeley, California, United States\\
$^{19}$ Department of Physics, University of Oslo, Oslo, Norway\\
$^{20}$ Department of Physics and Technology, University of Bergen, Bergen, Norway\\
$^{21}$ Dipartimento di Fisica, Universit\`{a} di Pavia, Pavia, Italy\\
$^{22}$ Dipartimento di Fisica dell'Universit\`{a} and Sezione INFN, Cagliari, Italy\\
$^{23}$ Dipartimento di Fisica dell'Universit\`{a} and Sezione INFN, Trieste, Italy\\
$^{24}$ Dipartimento di Fisica dell'Universit\`{a} and Sezione INFN, Turin, Italy\\
$^{25}$ Dipartimento di Fisica e Astronomia dell'Universit\`{a} and Sezione INFN, Bologna, Italy\\
$^{26}$ Dipartimento di Fisica e Astronomia dell'Universit\`{a} and Sezione INFN, Catania, Italy\\
$^{27}$ Dipartimento di Fisica e Astronomia dell'Universit\`{a} and Sezione INFN, Padova, Italy\\
$^{28}$ Dipartimento di Fisica `E.R.~Caianiello' dell'Universit\`{a} and Gruppo Collegato INFN, Salerno, Italy\\
$^{29}$ Dipartimento DISAT del Politecnico and Sezione INFN, Turin, Italy\\
$^{30}$ Dipartimento di Scienze MIFT, Universit\`{a} di Messina, Messina, Italy\\
$^{31}$ Dipartimento Interateneo di Fisica `M.~Merlin' and Sezione INFN, Bari, Italy\\
$^{32}$ European Organization for Nuclear Research (CERN), Geneva, Switzerland\\
$^{33}$ Faculty of Electrical Engineering, Mechanical Engineering and Naval Architecture, University of Split, Split, Croatia\\
$^{34}$ Faculty of Engineering and Science, Western Norway University of Applied Sciences, Bergen, Norway\\
$^{35}$ Faculty of Nuclear Sciences and Physical Engineering, Czech Technical University in Prague, Prague, Czech Republic\\
$^{36}$ Faculty of Physics, Sofia University, Sofia, Bulgaria\\
$^{37}$ Faculty of Science, P.J.~\v{S}af\'{a}rik University, Ko\v{s}ice, Slovak Republic\\
$^{38}$ Frankfurt Institute for Advanced Studies, Johann Wolfgang Goethe-Universit\"{a}t Frankfurt, Frankfurt, Germany\\
$^{39}$ Fudan University, Shanghai, China\\
$^{40}$ Gangneung-Wonju National University, Gangneung, Republic of Korea\\
$^{41}$ Gauhati University, Department of Physics, Guwahati, India\\
$^{42}$ Helmholtz-Institut f\"{u}r Strahlen- und Kernphysik, Rheinische Friedrich-Wilhelms-Universit\"{a}t Bonn, Bonn, Germany\\
$^{43}$ Helsinki Institute of Physics (HIP), Helsinki, Finland\\
$^{44}$ High Energy Physics Group,  Universidad Aut\'{o}noma de Puebla, Puebla, Mexico\\
$^{45}$ Horia Hulubei National Institute of Physics and Nuclear Engineering, Bucharest, Romania\\
$^{46}$ HUN-REN Wigner Research Centre for Physics, Budapest, Hungary\\
$^{47}$ Indian Institute of Technology Bombay (IIT), Mumbai, India\\
$^{48}$ Indian Institute of Technology Indore, Indore, India\\
$^{49}$ INFN, Laboratori Nazionali di Frascati, Frascati, Italy\\
$^{50}$ INFN, Sezione di Bari, Bari, Italy\\
$^{51}$ INFN, Sezione di Bologna, Bologna, Italy\\
$^{52}$ INFN, Sezione di Cagliari, Cagliari, Italy\\
$^{53}$ INFN, Sezione di Catania, Catania, Italy\\
$^{54}$ INFN, Sezione di Padova, Padova, Italy\\
$^{55}$ INFN, Sezione di Pavia, Pavia, Italy\\
$^{56}$ INFN, Sezione di Torino, Turin, Italy\\
$^{57}$ INFN, Sezione di Trieste, Trieste, Italy\\
$^{58}$ Inha University, Incheon, Republic of Korea\\
$^{59}$ Institute for Gravitational and Subatomic Physics (GRASP), Utrecht University/Nikhef, Utrecht, Netherlands\\
$^{60}$ Institute of Experimental Physics, Slovak Academy of Sciences, Ko\v{s}ice, Slovak Republic\\
$^{61}$ Institute of Physics, Homi Bhabha National Institute, Bhubaneswar, India\\
$^{62}$ Institute of Physics of the Czech Academy of Sciences, Prague, Czech Republic\\
$^{63}$ Institute of Space Science (ISS), Bucharest, Romania\\
$^{64}$ Institut f\"{u}r Kernphysik, Johann Wolfgang Goethe-Universit\"{a}t Frankfurt, Frankfurt, Germany\\
$^{65}$ Instituto de Ciencias Nucleares, Universidad Nacional Aut\'{o}noma de M\'{e}xico, Mexico City, Mexico\\
$^{66}$ Instituto de F\'{i}sica, Universidade Federal do Rio Grande do Sul (UFRGS), Porto Alegre, Brazil\\
$^{67}$ Instituto de F\'{\i}sica, Universidad Nacional Aut\'{o}noma de M\'{e}xico, Mexico City, Mexico\\
$^{68}$ iThemba LABS, National Research Foundation, Somerset West, South Africa\\
$^{69}$ Jeonbuk National University, Jeonju, Republic of Korea\\
$^{70}$ Johann-Wolfgang-Goethe Universit\"{a}t Frankfurt Institut f\"{u}r Informatik, Fachbereich Informatik und Mathematik, Frankfurt, Germany\\
$^{71}$ Korea Institute of Science and Technology Information, Daejeon, Republic of Korea\\
$^{72}$ KTO Karatay University, Konya, Turkey\\
$^{73}$ Laboratoire de Physique Subatomique et de Cosmologie, Universit\'{e} Grenoble-Alpes, CNRS-IN2P3, Grenoble, France\\
$^{74}$ Lawrence Berkeley National Laboratory, Berkeley, California, United States\\
$^{75}$ Lund University Department of Physics, Division of Particle Physics, Lund, Sweden\\
$^{76}$ Nagasaki Institute of Applied Science, Nagasaki, Japan\\
$^{77}$ Nara Women{'}s University (NWU), Nara, Japan\\
$^{78}$ National and Kapodistrian University of Athens, School of Science, Department of Physics , Athens, Greece\\
$^{79}$ National Centre for Nuclear Research, Warsaw, Poland\\
$^{80}$ National Institute of Science Education and Research, Homi Bhabha National Institute, Jatni, India\\
$^{81}$ National Nuclear Research Center, Baku, Azerbaijan\\
$^{82}$ National Research and Innovation Agency - BRIN, Jakarta, Indonesia\\
$^{83}$ Niels Bohr Institute, University of Copenhagen, Copenhagen, Denmark\\
$^{84}$ Nikhef, National institute for subatomic physics, Amsterdam, Netherlands\\
$^{85}$ Nuclear Physics Group, STFC Daresbury Laboratory, Daresbury, United Kingdom\\
$^{86}$ Nuclear Physics Institute of the Czech Academy of Sciences, Husinec-\v{R}e\v{z}, Czech Republic\\
$^{87}$ Oak Ridge National Laboratory, Oak Ridge, Tennessee, United States\\
$^{88}$ Ohio State University, Columbus, Ohio, United States\\
$^{89}$ Physics department, Faculty of science, University of Zagreb, Zagreb, Croatia\\
$^{90}$ Physics Department, Panjab University, Chandigarh, India\\
$^{91}$ Physics Department, University of Jammu, Jammu, India\\
$^{92}$ Physics Program and International Institute for Sustainability with Knotted Chiral Meta Matter (SKCM2), Hiroshima University, Hiroshima, Japan\\
$^{93}$ Physikalisches Institut, Eberhard-Karls-Universit\"{a}t T\"{u}bingen, T\"{u}bingen, Germany\\
$^{94}$ Physikalisches Institut, Ruprecht-Karls-Universit\"{a}t Heidelberg, Heidelberg, Germany\\
$^{95}$ Physik Department, Technische Universit\"{a}t M\"{u}nchen, Munich, Germany\\
$^{96}$ Politecnico di Bari and Sezione INFN, Bari, Italy\\
$^{97}$ Research Division and ExtreMe Matter Institute EMMI, GSI Helmholtzzentrum f\"ur Schwerionenforschung GmbH, Darmstadt, Germany\\
$^{98}$ Saga University, Saga, Japan\\
$^{99}$ Saha Institute of Nuclear Physics, Homi Bhabha National Institute, Kolkata, India\\
$^{100}$ School of Physics and Astronomy, University of Birmingham, Birmingham, United Kingdom\\
$^{101}$ Secci\'{o}n F\'{\i}sica, Departamento de Ciencias, Pontificia Universidad Cat\'{o}lica del Per\'{u}, Lima, Peru\\
$^{102}$ Stefan Meyer Institut f\"{u}r Subatomare Physik (SMI), Vienna, Austria\\
$^{103}$ SUBATECH, IMT Atlantique, Nantes Universit\'{e}, CNRS-IN2P3, Nantes, France\\
$^{104}$ Sungkyunkwan University, Suwon City, Republic of Korea\\
$^{105}$ Suranaree University of Technology, Nakhon Ratchasima, Thailand\\
$^{106}$ Technical University of Ko\v{s}ice, Ko\v{s}ice, Slovak Republic\\
$^{107}$ The Henryk Niewodniczanski Institute of Nuclear Physics, Polish Academy of Sciences, Cracow, Poland\\
$^{108}$ The University of Texas at Austin, Austin, Texas, United States\\
$^{109}$ Universidad Aut\'{o}noma de Sinaloa, Culiac\'{a}n, Mexico\\
$^{110}$ Universidade de S\~{a}o Paulo (USP), S\~{a}o Paulo, Brazil\\
$^{111}$ Universidade Estadual de Campinas (UNICAMP), Campinas, Brazil\\
$^{112}$ Universidade Federal do ABC, Santo Andre, Brazil\\
$^{113}$ Universitatea Nationala de Stiinta si Tehnologie Politehnica Bucuresti, Bucharest, Romania\\
$^{114}$ University of Cape Town, Cape Town, South Africa\\
$^{115}$ University of Houston, Houston, Texas, United States\\
$^{116}$ University of Jyv\"{a}skyl\"{a}, Jyv\"{a}skyl\"{a}, Finland\\
$^{117}$ University of Kansas, Lawrence, Kansas, United States\\
$^{118}$ University of Liverpool, Liverpool, United Kingdom\\
$^{119}$ University of Science and Technology of China, Hefei, China\\
$^{120}$ University of South-Eastern Norway, Kongsberg, Norway\\
$^{121}$ University of Tennessee, Knoxville, Tennessee, United States\\
$^{122}$ University of the Witwatersrand, Johannesburg, South Africa\\
$^{123}$ University of Tokyo, Tokyo, Japan\\
$^{124}$ University of Tsukuba, Tsukuba, Japan\\
$^{125}$ Universit\"{a}t M\"{u}nster, Institut f\"{u}r Kernphysik, M\"{u}nster, Germany\\
$^{126}$ Universit\'{e} Clermont Auvergne, CNRS/IN2P3, LPC, Clermont-Ferrand, France\\
$^{127}$ Universit\'{e} de Lyon, CNRS/IN2P3, Institut de Physique des 2 Infinis de Lyon, Lyon, France\\
$^{128}$ Universit\'{e} de Strasbourg, CNRS, IPHC UMR 7178, F-67000 Strasbourg, France, Strasbourg, France\\
$^{129}$ Universit\'{e} Paris-Saclay, Centre d'Etudes de Saclay (CEA), IRFU, D\'{e}partment de Physique Nucl\'{e}aire (DPhN), Saclay, France\\
$^{130}$ Universit\'{e}  Paris-Saclay, CNRS/IN2P3, IJCLab, Orsay, France\\
$^{131}$ Universit\`{a} degli Studi di Foggia, Foggia, Italy\\
$^{132}$ Universit\`{a} del Piemonte Orientale, Vercelli, Italy\\
$^{133}$ Universit\`{a} di Brescia, Brescia, Italy\\
$^{134}$ Variable Energy Cyclotron Centre, Homi Bhabha National Institute, Kolkata, India\\
$^{135}$ Warsaw University of Technology, Warsaw, Poland\\
$^{136}$ Wayne State University, Detroit, Michigan, United States\\
$^{137}$ Yale University, New Haven, Connecticut, United States\\
$^{138}$ Yonsei University, Seoul, Republic of Korea\\
$^{139}$  Zentrum  f\"{u}r Technologie und Transfer (ZTT), Worms, Germany\\
$^{140}$ Affiliated with an institute covered by a cooperation agreement with CERN\\
$^{141}$ Affiliated with an international laboratory covered by a cooperation agreement with CERN.\\

\end{flushleft} 
\end{document}